\documentclass[aps,prx,twocolumn,notitlepage,amsmath,amstex,amssymb,superscriptaddress]{revtex4-1}

\usepackage{lmodern} % readable font in pdf
\usepackage{microtype} % better font spacing
\usepackage[english]{babel}

\usepackage{color,graphicx}

\usepackage{dsfont,url}
\usepackage{slashed,cancel}
\usepackage[usenames,dvipsnames]{xcolor}
\usepackage[colorlinks=true, linkcolor=black, citecolor=ForestGreen]{hyperref}%BrickRed
\usepackage{mdframed}

\usepackage{physics,tikz}

\usepackage[caption=false]{subfig}

\usetikzlibrary{decorations.markings,calc}

\begin{document}

\title{\bf 
Hydrodynamic tails in chaotic spin chains with quantum group symmetry
}

\author{Luca V. Delacr\'etaz}%\email{asd}
\thanks{Authors listed alphabetically}
\affiliation{Leinweber Institute for Theoretical Physics \& James Franck Institute, University of Chicago, Chicago, IL 60637, USA}
\author{Victor Gorbenko}
\affiliation{Laboratory for Theoretical Fundamental Physics, Institute of Physics,\\ École Polytechnique Fédérale de Lausanne, CH-1015 Lausanne, Switzerland}
\author{Jiaozi Wang}
\affiliation{Institute of Fundamental Physics and Quantum Technology, and
School of Physical Science and Technology,\\ Ningbo University, Ningbo, Zhejiang 315211, China}

\author{Bernardo Zan}
\affiliation{Dipartimento di Fisica, Università di Genova and INFN, Sezione di Genova,\\ Via Dodecaneso 33, 16146, Genoa, Italy}
\author{Aleksandr Zhabin}
\affiliation{Laboratory for Theoretical Fundamental Physics, Institute of Physics,\\ École Polytechnique Fédérale de Lausanne, CH-1015 Lausanne, Switzerland}

\begin{abstract}
The interplay between symmetry and thermalization governs the late-time dynamics of local quantum and classical many-body systems at nonzero temperature. Recently, two parallel frontiers have emerged: the search for robust anomalous hydrodynamics---such as superdiffusion---in generic, non-integrable models, and the formal effort to generalize the fundamental concept of global symmetry. In this paper, we bridge these frontiers by demonstrating that quantum group symmetry provides a novel mechanism for anomalous hydrodynamics in chaotic systems. We study the dynamics of local operators carrying $U(1)$ charge in non-integrable lattice models that also have quantum group symmetry. One example is transverse spin in the XXZ model with integrability breaking deformations. While such excitations are expected to decay very quickly at high temperature because their charge forbids overlap with conventional hydrodynamic densities, we find that protection by the quantum group symmetry makes these modes long-lived, despite the absence of local quantum group charge density or current. Furthermore, the dynamics is superdiffusive across Hamiltonian, Floquet, and classical realizations, and exhibits unusual finite size effects at very late times. We also revisit transverse spin dynamics in the integrable XXZ model.
\end{abstract}

%~ \date{}
\maketitle

%######################################################################%
%======================================================================%
%======================================================================%
%======================================================================%
%######################################################################%
\section{Introduction}

Symmetry guides physical theories across the sciences. In the past decade, the concept of symmetry in many-body physics has been thoroughly reinvestigated and extended. It now  includes higher-form symmetries that can capture the conservation of extended objects such as flux lines, and categorical symmetries that do not have a group-like structure. 
However, most recent discoveries are generalizations of discrete symmetries.
{\em Continuous} symmetries instead have the most important consequences on local dynamics: they lead to local current conservation or Ward identities, produce Nambu-Goldstone bosons when spontaneously broken, and protect long-lived hydrodynamic excitations in thermal states. Finding generalizations of continuous symmetry beyond those classified by group theory could impact fields ranging from particle physics to quantum dynamics.

\begin{figure}
\includegraphics[width=0.85\columnwidth]{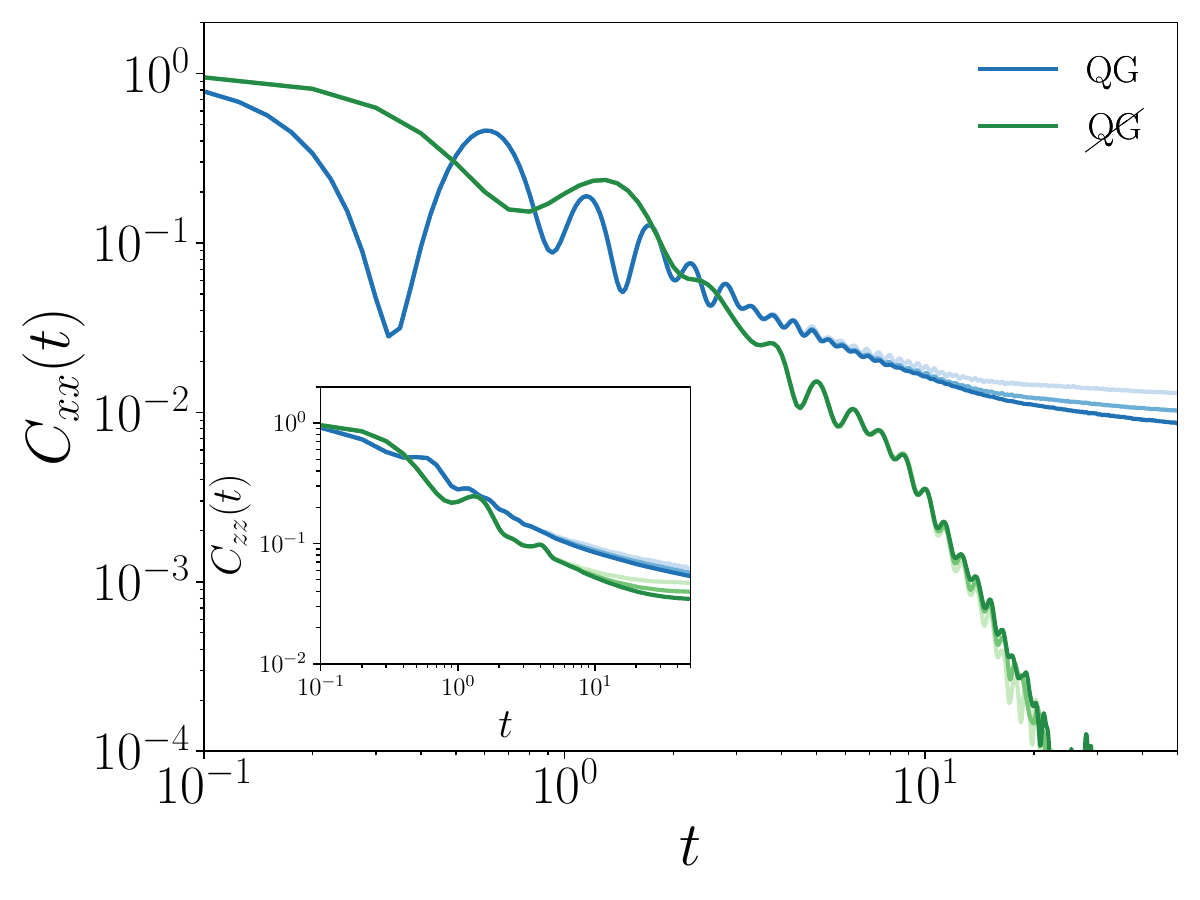}
\caption{Spin autocorrelation function $C_{ab}(t)=\langle \sigma^a_i(t)\sigma^b_i\rangle$ in chaotic XXZ-like models with and without quantum group (QG) symmetry, for system sizes $L = 22,26,30$. Both models feature diffusive decay of the longitudinal spin $\sigma_z$, the conserved density for a conventional $U(1)$ symmetry (inset). In contrast, the decay of transverse spin is markedly different in both models: it is faster than polynomial in the model without QG symmetry, as expected for a non-conserved operator, but features slow polynomial decay in the QG symmetric model.}
\label{fig_money}
\end{figure}

In this paper, we test whether quantum groups usefully generalize continuous global symmetry. Specifically, we study whether such a structure can protect long-lived hydrodynamic modes, by numerically investigating non-integrable unitary lattice models with quantum group symmetry. One class of models that we consider is the one-dimensional XXZ model with next-nearest neighbor deformations
\begin{equation}\label{eq_XXZ}
H = \sum_i \sigma^x_i \sigma^x_{i+1} + \sigma^y_i \sigma^y_{i+1} + \Delta \sigma^z_i \sigma^z_{i+1} + V_i\, .
\end{equation}
Deformations $V\neq0$ generically break the integrability of the XXZ model and lead to chaotic thermalizing dynamics. If they preserve the $U(1)$ spin rotation symmetry around the $z$-axis, the longitudinal spin $\sigma^z$ is expected to diffuse at late times because it is a conserved density: $\langle\sigma^z(t)\sigma^z\rangle \sim 1/\sqrt{t}$. If the full $SU(2)$ symmetry of the $\Delta=1$ model is preserved, this also applies to transverse spin $\sigma^x,\,\sigma^y$. In contrast, when $\Delta\neq 1$ the transverse spin is not conserved and is expected to decay quickly (faster than any power-law) in chaotic models.

Strikingly, we find that a certain class of next-nearest neighbor interactions $V$ feature slow power-law decay of transverse spin, even though they break both integrability and $SU(2)$ down to $U(1)$, see Fig.~\ref{fig_money}. The deformations that lead to this behavior are special in that they preserve a quantum-deformed $SU(2)$ symmetry. Our results show that even though quantum groups do not provide local conserved currents, they can protect long-lived hydrodynamic modes and lead to slow decay of local operators.

Another surprising feature of the decay of transverse spin in these models is that it appears to be superdiffusive. While a theory of fluctuating hydrodynamics for quantum group charges is beyond the scope of this work, these results suggest that such a construction should allow for rich anomalous hydrodynamics. It would be interesting to revisit other models showing apparent anomalous hydrodynamics through the lens of quantum group and other generalized symmetries. 

We start by confirming that the Hamiltonians with next nearest neighbor deformations preserving quantum group symmetry are chaotic in Sec.~\ref{sec_RMT}. Transverse spin dynamics is then studied in these models in Sec.~\ref{sec_main}, as well as in related Floquet and classical models, confirming slow power-law decay across models with a shared superdiffusive universality class. Spin chains with quantum group symmetry also feature peculiar late time saturation of correlation functions in finite systems, which are discussed in Sec.~\ref{sec_finite_size}. Some features of quantum group symmetry in chaotic spin chains also carry over to integrable spin chains such as the XXZ model, which we explore in Sec.~\ref{sec_integrable}. We conclude in Sec.~\ref{sec_discuss}.

%######################################################################%
%======================================================================%
%======================================================================%
%======================================================================%
%######################################################################%
\section{Quantum chaos in local unitary models with QG symmetry}\label{sec_RMT}

Quantum group (QG) symmetry was originally discovered in the study of integrable spin-chains \cite{Takhtajan:1979iv,Kulish:1981dli,Drinfeld:1986in}, and has thus often been associated with integrability. However, QG symmetry does not require integrability, and can act as a global continuous symmetry group, an angle recently emphasized in  \cite{Gabai:2024puk,Gabai:2024qum, Gorbenko:2025wzs}. 

We focus on the simplest QG symmetry, $U_q(sl_2)$, which is a $q$-deformation of $SU(2)$ global symmetry. In principle, $q$ can be any complex number, but in order to make the model unitary we will consider $q \in \mathbb{R}$. A starting point of our construction is the XXZ Hamiltonian \eqref{eq_XXZ} in the easy-axis regime $\Delta = \frac12 (q+q^{-1})\geq 1$, with a specific boundary term
\begin{align}\label{eq:h0}
    H_0 = \sum_{i=1}^{L} \left(\sigma^x_i \sigma^x_{i+1} + \sigma^y_i \sigma^y_{i+1} + \tfrac12(q+q^{-1}) \sigma^z_i \sigma^z_{i+1} \right) \notag\\
    -\tfrac12(q-q^{-1}) \left(\sigma^z_1 - \sigma^z_L\right)\, .
\end{align}
$H_0$ is known to have $U_q(sl_2)$ symmetry \cite{Pasquier:1989kd}. In the limit $q\to 1$ it reduces to the open XXX chain with $SU(2)$ symmetry. $H_0$ has QG symmetry because it is constructed from a linear combination of Hecke algebra elements,
\begin{align}\label{eq_Hecke}
    R_{i}^{\rm QG} = \tfrac12\Big[\sigma_{i}^{x} \sigma_{i+1}^{x} + \sigma_{i}^{y} \sigma_{i+1}^{y} + \tfrac{q+q^{-1}}2 \big( \sigma_{i}^{z}\sigma_{i+1}^{z} + 1 \big)  \notag\\ 
    -\tfrac{q-q^{-1}}2 \big( \sigma_{i}^{z} - \sigma_{i+1}^{z} - 2 \big) \Big]\,,
\end{align}
which commute with the $U_q(sl_2)$ generators on the Hilbert space of a spin chain, see \cite{Jones:1987dy,Gorbenko:2025wzs} for more details.

The integrability of the XXZ Hamiltonian \eqref{eq:h0} can be broken, e.g., by adding a next-nearest neighbor interaction
\begin{align}\label{eq:standard_nnn}
V_i^{\rm nnn}
	&= \sigma_i^x\sigma_{i+2}^x + \sigma_i^y\sigma_{i+2}^y + \sigma_i^z\sigma_{i+2}^z\, .
\end{align}
While this deformation also breaks the QG symmetry, a specific combination of three-site terms preserves it:
\begin{align}\label{def:Vi}
V_i^{\rm QG}
	&= V_i^{\rm nnn} + A_i + B_i\, ,
\end{align}
where the terms $A_i$ and $B_i$ are given by
\begin{align}
A_i
	&= 
	%~ \frac{(q-q^{-1})^2}{4}
	\tfrac1{4}(q-q^{-1})^2
	\left(\sigma_i^z \sigma_{i+1}^z + \sigma_{i+1}^z \sigma_{i+2}^z\right)\notag\\ 
	&\quad- \tfrac14(q^2 - q^{-2})\left(\sigma_i^z - \sigma_{i+2}^z\right)\, \\
B_i
	&= \tfrac12(q-q^{-1})\bigl(\sigma_i^x \sigma_{i+1}^x\sigma_{i+2}^z+\sigma_i^y \sigma_{i+1}^y\sigma_{i+2}^z \notag\\ 
	&\qquad\qquad\qquad - \sigma_i^z \sigma_{i+1}^x\sigma_{i+2}^x-\sigma_i^z \sigma_{i+1}^y\sigma_{i+2}^y  \bigr)\, .
\end{align}
One can show that the combination $V_i^{\rm QG}$ is QG invariant because it is made out of Hecke algebra elements \cite{Gorbenko:2025wzs}. Thus, the symmetry is preserved by any Hamiltonian of the form $\sum_i \alpha_i V_i^{\rm QG}$ with $\alpha_i\in \mathbb R$. We consider two such Hamiltonians that also have lattice translation symmetry:
\begin{align}\label{eq_HQG}
H_{\rm QG}
	&= H_0 + \lambda\sum_{i=1}^{L}\  V_i^{\rm QG}\, ,\\
H_{\rm QG, \, stag} \label{eq_Hstag}
	&= H_0 + \lambda\sum_{i=1}^{L}(-1)^i V_i^{\rm QG}\, ,
\end{align}
We emphasize that we only consider unitary models, with $q\in \mathbb R$ \footnote{The first of these two Hamiltonians, albeit with $|q|=1$ and different boundary conditions, was studied in \cite{Ikhlef_2009}.}. \phantom{\cite{Ikhlef_2009}}
% In the integrable regime ($\lambda=0$), these reduce to the XXZ model in the easy-axis regime $\Delta  = \frac12 (q+q^{-1})\geq 1$. Changing $\lambda>0$ introduces integrability breaking. 

\begin{figure}[t]
\includegraphics[width=1.0\columnwidth]{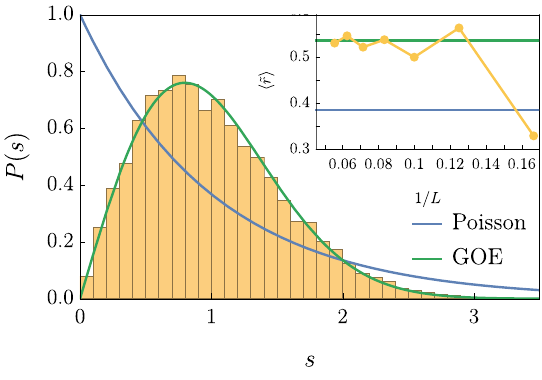}
\caption{ Level spacing distribution, $P(s)$, for the Hamiltonian $H_{\rm QG, \, stag}$ with $L=18$, $q=2.57$, $\lambda=1$. Inset: convergence of average level spacing ratio, $\langle\Tilde{r}\rangle$, as a function of $1/L$. Blue and green lines are the Poisson and RMT predictions respectively . 
}
\label{fig_Ps}
\end{figure}

We now confirm that these QG symmetric models are chaotic, with eigenvalue statistics described by random matrix theory (RMT). Both Hamiltonians are real orthogonal matrices,
\begin{align}
    H_{\rm QG} = H_{\rm QG}^{T}, \qquad H_{\rm QG, \, stag} = H_{\rm QG, \, stag}^{T}\,.
\end{align}
We will thus compare their eigenvalue statistics to the Gaussian Orthogonal Ensemble (GOE) after accounting for their symmetries. The QG symmetry can be dealt with as follows \cite{Gorbenko:2025wzs}: the states in $U_q(sl_2)$ representations are labeled by spin and its projection $\{\ell,m\}$, where $m \in \{-\ell, \ldots, \ell\}$. In order to study eigenvalues in a sector with fixed quantum numbers $\{\ell,m\}$, one can diagonalize the Hamiltonians in the sectors with $S^z=m$ and $S^z=m+1$, and then remove from the first set of eigenvalues all those that are also contained in the second set. 

In addition to $U_q(sl_2)$, $H_{\rm QG}$ has an additional $q$-deformed reflection (or parity) symmetry, which becomes conventional spatial reflection symmetry for $q\to1$. In order to analyze the spectral statistics of eigenvalues one has to project onto an irreducible subsector where all the symmetries are resolved. However, some of the eigenvalues of $q$-parity operator scale as $\sim q^{L^2/4}$, which makes the procedure numerically difficult unless $q$ is very close to unity \cite{Gorbenko:2025wzs}. For this reason, it is much simpler to establish the RMT behavior of the staggered model \eqref{eq_Hstag}, for which this symmetry is absent. Fig.~\ref{fig_Ps} shows that its nearest level spacing statistics $s_i=E_{i+1}-E_i$ agrees with the GOE prediction $P(s) = \frac{\pi s}{2} e^{-\frac{\pi s^2}{4}}$. Low- and high-energy tails of the spectrum were removed, followed by the local unfolding procedure. Another convenient metric to study how close the distribution is to RMT predictions is the average value of the ratio between consecutive level spacings,
$\Tilde{r}_i = \min \left( \frac{s_i}{s_{i+1}}, \frac{s_{i+1}}{s_i} \right)$. The advantage of this quantity is that no local unfolding procedure is required. The value of the average ratio for Poisson and Wigner distributions are $\langle\Tilde{r}\rangle_{\rm P} \simeq 0.386$ and $\langle\Tilde{r}\rangle_{\rm GOE} \simeq 0.536$. Fig.~\ref{fig_Ps} shows that $\langle\Tilde{r}\rangle$ approaches the GOE value as system size is increased.

\begin{figure}[t]
\includegraphics[width=1.0\columnwidth]{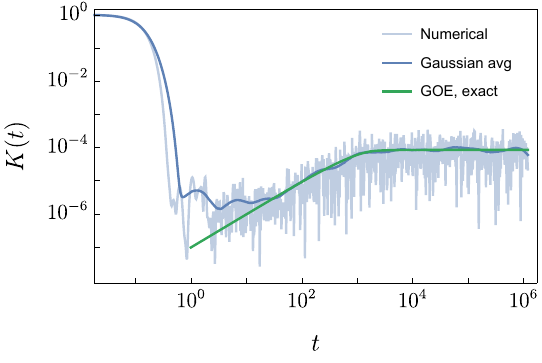}
\caption{ Spectral form factor $K(t)$ for the Hamiltonian $H_{\rm QG\, stag}$ with $L=18$, $q=2.57$, $\lambda=1$ in the sector $\{\ell=1,m=1\}$ of size $N=11934$. The raw data (light blue) is Gaussian averaged (dark blue) to remove noise and compare to the RMT prediction from Eq.~\eqref{eq:SFF_GOE} (green).
}
\label{fig_SFF}
\end{figure}

A more refined probe of quantum chaos that connects to the time-dependent correlation functions studied in the next sections is the spectral form factor (SFF)
\begin{align}
    K(t) = \frac{1}{N^2} \sum_{i,j} e^{i(E_i-E_j)t}\,,
\end{align}
where $E_i$ are the energies of the Hamiltonian and $N$ is the dimensionality of the sector. The SFF at different times captures spectral correlations across different energy separations. The RMT prediction features a distinctive ramp and plateau structure, which for GOE take the form:
\begin{align}\label{eq:SFF_GOE}
    K_{\text{GOE}}(t) = \frac{1}{N} \begin{cases} \dfrac{2t}{t_H} - \dfrac{t}{t_H}\log\left(1 + \dfrac{2t}{t_H}\right) & t \leq t_H \\
    2 - \dfrac{t}{t_H}\log\left(\dfrac{\frac{2t}{t_H}+1}{\frac{2t}{t_H}-1}\right) & t > t_H \end{cases}
\end{align}
In this expression the Heisenberg time is defined as $t_H = 2\pi/\Delta E$, where $\Delta E$ is the mean level spacing. Fig.~\ref{fig_SFF} shows agreement of the averaged SFF with the RMT prediction.

%######################################################################%
%======================================================================%
%======================================================================%
%======================================================================%
%######################################################################%
\section{Power-law decay of transverse spin}\label{sec_main}

%======================================================================%
%======================================================================%
%======================================================================%
\subsection{Autocorrelation function}\label{sse_autocorrelation}

In chaotic versions of the XXZ spin chain, the transverse spin $\sigma^+ = \sigma^x + i \sigma^y$ is an example of an operator `orthogonal' to hydrodynamics: since it carries a $U(1)$ charge, no composite object made out of hydrodynamic densities---which are neutral---can match its quantum numbers. It is therefore expected to decay quickly in thermal states; we confirm this expectation in Fig.~\ref{fig_money}.

\begin{figure}[t]
\includegraphics[width=1\columnwidth]{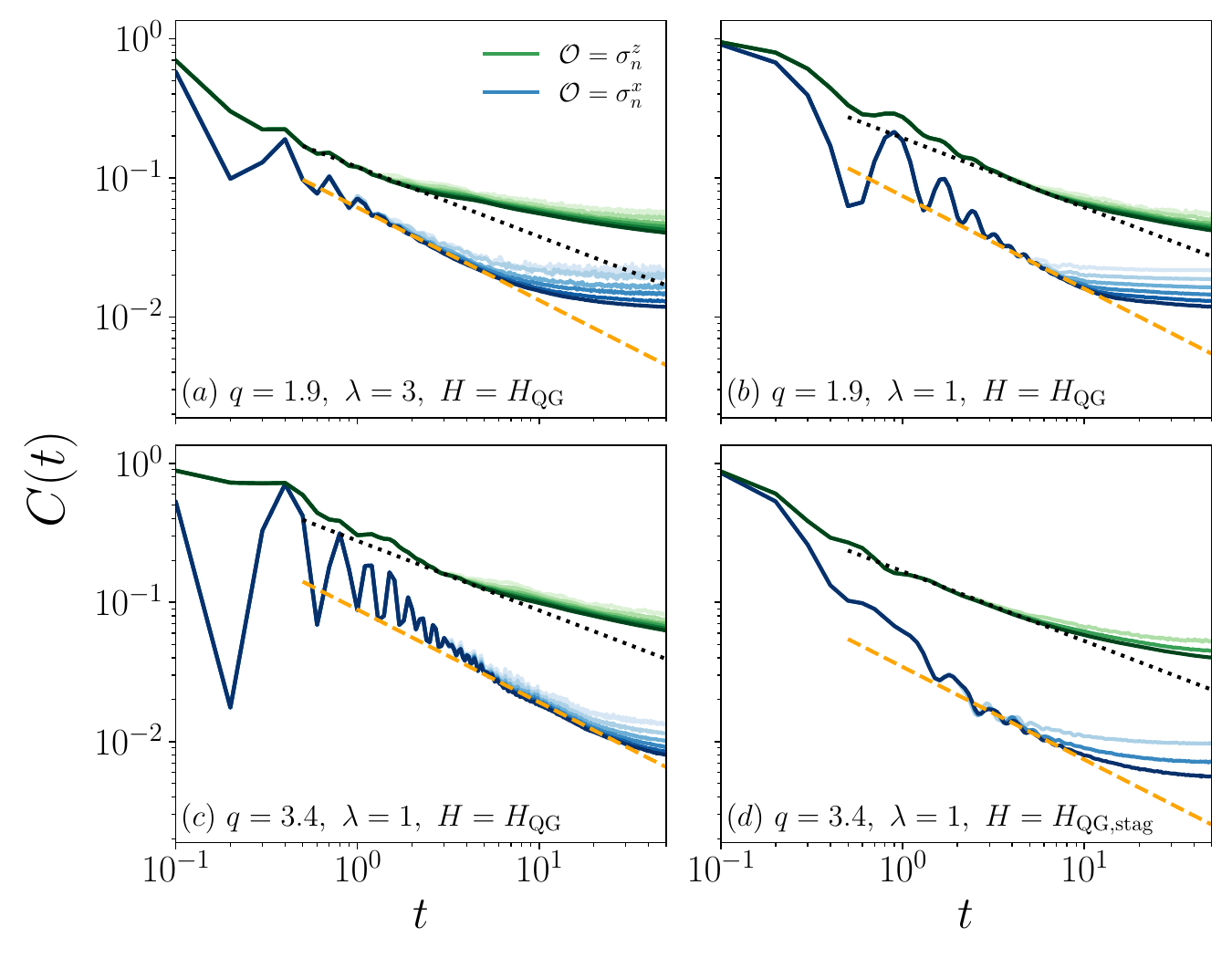}
\caption{Infinite temperature autocorrelation function $C(t)= \Tr(\mathcal O(t)\mathcal O)$ for transverse and longitudinal spin $\mathcal O = \sigma^x,\,\sigma^z$ in QG symmetric Hamiltonian models \eqref{eq_HQG} and \eqref{eq_Hstag} with various parameters. The light to dark colors correspond to increasing system size: $L = 20,22,24,26,28,30$ in (a,b,c) and $L = 22,26,30$ in (d).
Dotted ($1/t^{1/2}$) and dashed ($1/t^{2/3}$) lines are guides to the eye.}
\label{fig_C_XXt_Hamiltonian}
\end{figure}

As anticipated in Fig.~\ref{fig_money}, the chaotic spin chains with QG symmetry introduced in Sec.~\ref{sec_RMT} have an entirely different behavior, showing slow power-law decay in the thermodynamic limit. Fig.~\ref{fig_C_XXt_Hamiltonian} confirms this observation for both Hamiltonian models \eqref{eq_HQG} and \eqref{eq_Hstag} with several different choices of parameters, studied numerically using dynamical quantum typicality. Transverse spin appears to be superdiffusive, $\langle\sigma^+_i(t)\sigma^-_i\rangle \sim 1/t^{1/z}$, with an exponent $z\approx 1.6$  that is largely insensitive to parameters. Longitudinal spin instead appears to exhibit conventional diffusion $\langle \sigma^z_i(t)\sigma^z_i\rangle\sim 1/\sqrt{t}$ for all cases, as expected for chaotic models with a global $U(1)$ symmetry. 

\begin{figure}[t]
\includegraphics[width=1\columnwidth]{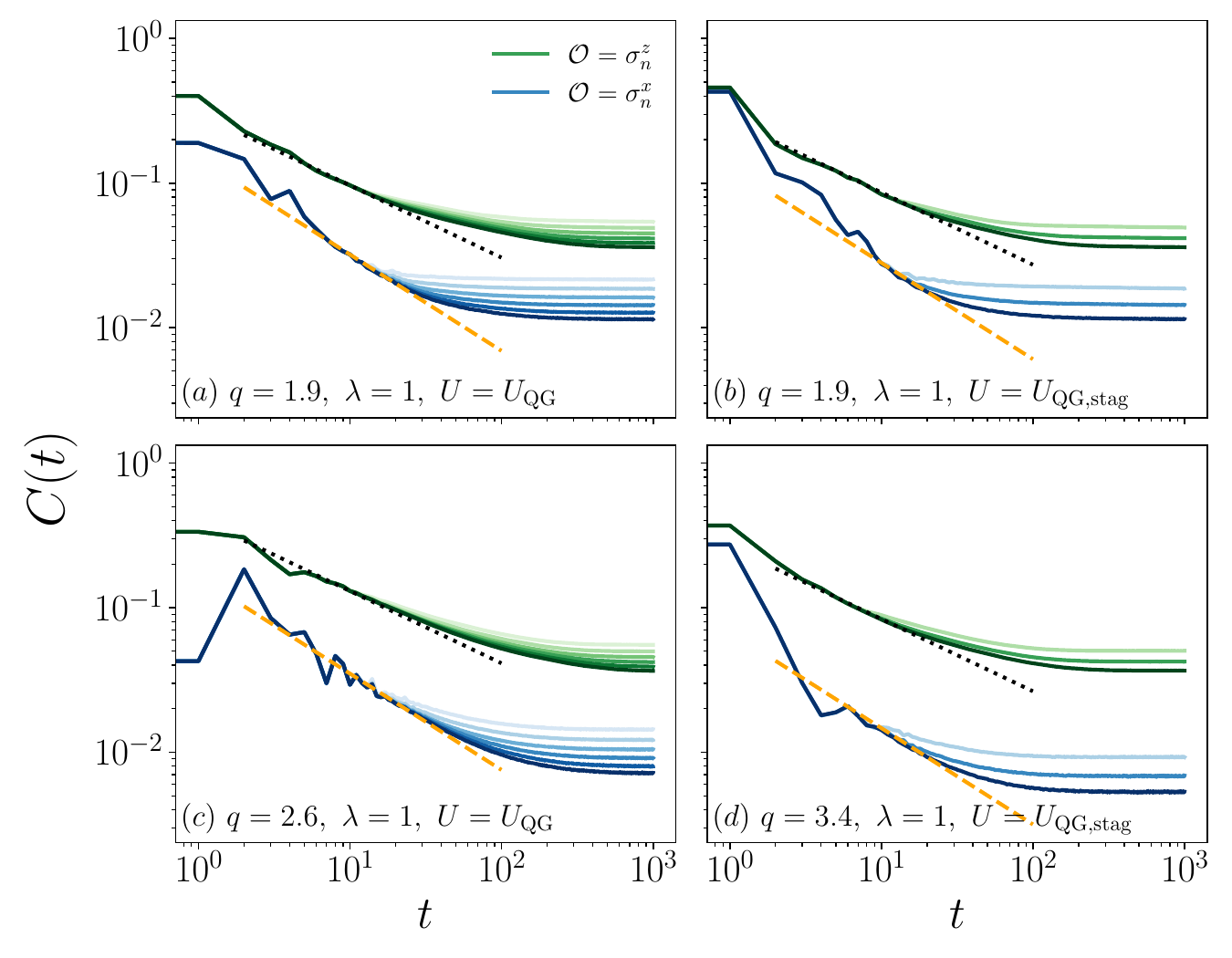}
\caption{Similar to Fig.~\ref{fig_C_XXt_Hamiltonian}, but for Floquet models with QG symmetry described in Eq.~\eqref{eq_Floquet}.
System sizes are $L=20,22,24,26,28,30$ in panels (a) and (c), and $L=22,26,30$ in panels (b) and (d), with colors from light to dark as $L$ increases.
}
\label{fig_C_XXt_Floquet}
\end{figure}

To explore the scope of this unusual behavior, we also consider Floquet models with QG symmetry. The results are shown in Fig.~\ref{fig_C_XXt_Floquet}, and are very similar to the Hamiltonian system: we still observe slow power-law decay of transverse spin despite the absence of conventional global symmetries protecting its conservation. The Floquet model is defined by
\begin{equation}\label{eq_Floquet}
U_{\mathrm{QG}} = U_{\lambda} U_{0},
\end{equation}
where, up to the detailed gate ordering specified in Appendix~\ref{app:floquet_map},
\begin{equation}
U_{0}
=
\prod_{\ell=1}^{L/2} e^{-i\tau R^{\mathrm{QG}}_{2\ell}}
\prod_{\ell=1}^{L/2} e^{-i\tau R^{\mathrm{QG}}_{2\ell-1}},
\end{equation}
implements the XXZ-like dynamics with $R_i^{\rm QG}$ defined in Eq.~\eqref{eq_Hecke}, and
\begin{equation}\label{def-Ulambda}
U_{\lambda}
=
\prod_{\ell=1}^{L/3} e^{-i\tau\lambda V_{3\ell}^{\mathrm{QG}}}  
\prod_{\ell=0}^{L/3} e^{-i\tau\lambda V_{3\ell-1}^{\mathrm{QG}}}
\prod_{\ell=1}^{L/3} e^{-i\tau\lambda V_{3\ell-2}^{\mathrm{QG}}},
\end{equation}
implements the integrability breaking deformation, with $V_{i}^{\mathrm{QG}}$ given in Eq.~\eqref{def:Vi}. The QG symmetry of these models follows from the fact that both local operators $R_i^{\rm QG}$ and $V_i^{\rm QG}$ are QG singlets.
Furthermore, we also consider a staggered version of the Floquet model, with the replacement $V^{\rm QG}_i \to (-1)^i V^{\rm QG}_i $ in $U_\lambda$ above. We set $ \tau = {\pi}/{8}$ throughout the paper.

It is interesting to contrast the unusual power-law decay of transverse spin we observe with other known behavior in nearby models. When $q=1$, our models reduce to the $SU(2)$ XXX model with an integrability breaking next-nearest neighbor deformation. The conventional global $SU(2)$ symmetry protects all spin components, which are expected to diffuse \cite{Dupont:2019iul,Glorioso:2020loc}. We confirm this behavior in App.~\ref{app_stretchedexp}. This conventional $SU(2)$ symmetry, which sets $\langle \sigma_i^x(t)\sigma_i^x\rangle = \langle \sigma_i^z(t)\sigma_i^z\rangle$, is clearly strongly broken in our QG models shown in Figs.~\ref{fig_C_XXt_Hamiltonian} and \ref{fig_C_XXt_Floquet}.

When $\lambda=0$, our models reduce to the integrable XXZ model in the `easy-axis' regime $\Delta  = \frac12 (q+q^{-1})\geq 1$. Longitudinal spin is known to decay diffusively when $\Delta >1$ (see, e.g., \cite{Znidaric:2011qvy}). Correlation functions of transverse spin instead have received surprisingly little attention in this otherwise intensively studied model, and we are not aware of any numerical or analytical results on this correlator at high temperature for $\Delta > 1$. More is known in other regimes: when $\Delta=0$ (XX model), the decay has long been known to be Gaussian $\sim e^{-at^2}$ \cite{brandt1976exact,capel1977autocorrelation} (see also \cite{DelVecchioDelVecchio:2021hjk}). It is believed to be exponential $e^{-\Gamma t}$ for $0<\Delta < 1$ \cite{PhysRevB.49.15669}, and in the KPZ universality class ($1/t^{3/2}$) at the isotropic point $\Delta =1$ \cite{Ljubotina:2019odz}. We study decay of transverse spin in the XXZ model with $\Delta>1$ in Sec.~\ref{sec_integrable}, and surprisingly find it to be close to {\em diffusive} ($1/t^{1/2}$). We do not believe that proximity to integrable models is what is responsible for the peculiar behavior observed in the models with QG symmetry. First, note that we consider large integrability breaking deformations $\lambda = O(1)$, and eigenvalue statistics agrees with RMT, as shown in Sec.~\ref{sec_RMT}. Second, the power-laws observed at intermediate time appear to be distinct. Finally the finite size effects observed in the autocorrelation function are strikingly different in both models, as will be discussed in Sec.~\ref{sec_finite_size}. However, we expect that the consequences of QG symmetry which we study include the decay of transverse spin in the XXZ model with $\Delta >1$, where it folds into a larger integrable structure. This is explored in Sec.~\ref{sec_integrable}.

We close this section by returning to chaotic versions of the XXZ model without QG symmetry. As shown already in Fig.~\ref{fig_money}, the transverse spin autocorrelation function is faster than any power-law. While it is often assumed that operators orthogonal to any hydrodynamic mode have exponentially decaying correlation function, we find that the decay is somewhat slower and appears to be a stretched exponential $\sim \exp(-t^\alpha)$, with $0<\alpha<1$, see App.~\ref{app_stretchedexp}. Similar behavior in different models was discussed in \cite{McCulloch:2025fzk,McCulloch:2026pra}. This interesting behavior warrants further investigation, but because it is not related to QG symmetry we do not do so here.

%======================================================================%
%======================================================================%
%======================================================================%
\subsection{Universality class and scaling function}

The precise determination of the dynamic critical exponent $z$ of a quantum many-body system is challenging due to the generic presence of power-law corrections to transport, which can be large at the intermediate times accessible numerically. In particular, Ref.~\cite{Michailidis:2023mkd} showed that diffusive systems with one conservation law generically appear to be superdiffusive at intermediate times, due to a non-negative correction which is only $1/\sqrt{t}$ suppressed in one spatial dimension: $\langle n(t)n\rangle \sim \frac1{\sqrt{t}} (1 + \frac{\alpha}{\sqrt{t}}+ \cdots)$, $\alpha \geq 0$. These corrections are large in the classical Heisenberg model \cite{Glorioso:2020loc}, which is not so distant from the models we consider here. It is therefore important to rule out the possibility that the apparent superdiffusive behavior observed in Figs.~\ref{fig_C_XXt_Hamiltonian} and \ref{fig_C_XXt_Floquet} is an intermediate time effect. There is a simple way to do so, thanks to the fact that the effective field theory of diffusion relates the coefficient $\alpha$ above to the density 3-point function \cite{Delacretaz:2023ypv}. In App.~\ref{app_ruleoutdiff}, we measure this 3-point function and show that the magnitude of $\alpha$ is far too small to produce a sizeable $1/\sqrt{t}$ correction to the autocorrelation function. We believe this rules out conventional diffusion of transverse spin in this model.

One way to further pinpoint the dissipative universality class of the QG models is to study position-resolved data. While this is a costly procedure, it gives access to the entire universal scaling function
\begin{equation}\label{eq_Ctx}
\langle \sigma^+_i(t) \sigma^-_j(0)\rangle \simeq \frac{1}{t^{1/z}} F \left(x/t^{1/z}\right)\, , \qquad x = i-j\, .
\end{equation}
This equation receives scaling corrections discussed above that are important at early times, as well as finite size corrections: it holds approximately between the local and global equilibration times. This strategy was useful in identifying the KPZ universality class in the integrable XXZ model at the isotropic point $\Delta=1$ \cite{Ljubotina:2019odz}.

\begin{figure}[t]
\includegraphics[width=1.0\columnwidth]{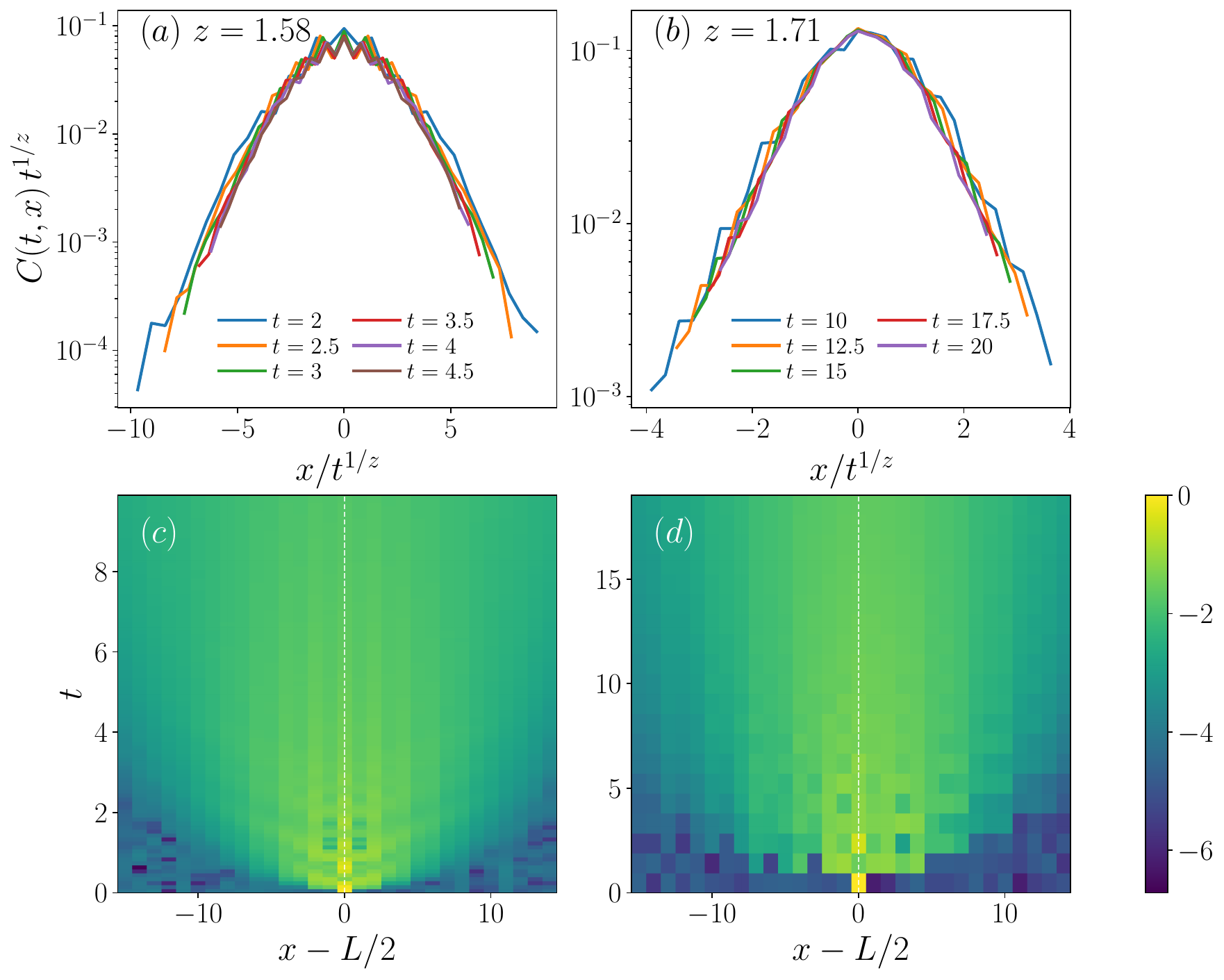}
\caption{Scaling collapse of $C(t,x)t^{1/z}$ versus $x/t^{1/z}$ for
transverse spin ${\cal O} = \sigma^x$ in
the (a) Hamiltonian model and (b) Floquet model for $q = 2.6$ and $\lambda = 1.0$. The corresponding heat maps of $C(t,x)$ in the $(x,t)$ plane are shown in (c) and (d), respectively. The system size is $L = 30$.}
\label{fig_collapse}
\end{figure}

Fig.~\ref{fig_collapse} shows the spacetime resolved correlator of transverse spin \eqref{eq_Ctx}. We observe good collapse, with a superdiffusive exponent $z\approx 1.6$-1.7. Furthermore, the scaling function appears to be the same for Hamiltonian and Floquet model, suggesting they are both described by the same dissipative universality class. This scaling function also appears to be distinct from that of diffusion ($F(y) = e^{-y^2}$) or KPZ. The scaling function for longitudinal spin is discussed in App.~\ref{app_data}.

The fact that densities associated with QG symmetry are superdiffusive $z<2$ is particularly interesting from the perspective of fluctuating hydrodynamics, as it suggests that the collective hydrodynamic excitations are strongly fluctuating. One well-known mechanism for superdiffusion in chaotic systems is sound modes in one spatial dimension, which are described by an equation of the form
\begin{equation}\label{eq_Burgers}
\partial_t n + c(n) \partial_x n + D \partial_x^2 n  + \cdots = 0\, ,
\end{equation}
where the ellipsis denotes noise and higher derivative terms. Nonlinearities in the velocity of sound modes $dc(n)/dn\neq 0$ are relevant in $d=1$, 
and lead to dissipation in the KPZ universality class for this equation (the noisy Burger's equation) \cite{PhysRevA.16.732}. An equation of this form describes fluctuating hydrodynamics in chaotic 1d systems with momentum conservation \cite{PhysRevLett.89.200601,Spohn2014,Delacretaz:2021ufg} (where Eq.~\eqref{eq_Burgers} really involves several conserved densities, corresponding to momentum and energy or charge conservation), as well as charge dynamics with a $U(1)$ chiral anomaly \cite{Delacretaz:2020jis} (as occurs on quantum Hall edges). A similar equation for two of the infinite hydrodynamic modes of the integrable and isotropic XXZ model can be used to justify the appearance of KPZ universality in that context \cite{DeNardis:2022dxm}. In chaotic models with QG symmetry, this equation cannot apply to transverse spin density $\sigma^\pm$ alone, as the $U(1)$ symmetry under which the density has unit charge would forbid the leading nonlinearity in Eq.~\eqref{eq_Burgers} that drives the system to KPZ. Furthermore, the fact that the scaling function observed in \eqref{fig_collapse} differs from that of KPZ suggests that our models exhibit a new dissipative universality class. One possible nonlinear equation for transverse spin is
\begin{equation}
\partial_t \sigma^+ + a(q) \sigma^+\partial_x n + D \partial_x^2 \sigma^+ + \cdots = 0\, ,
\end{equation}
with $a(1)=0$ to recover transverse spin diffusion when $q=1$. This convective nonlinearity mixing transverse spin to the $U(1)$ density $n=\sigma^z$ is likely to lead to superdiffusion. The presence of this type of nonlinear terms breaking $\sigma^+$ conservation also seems natural given the form the nonlocal QG density discussed later (see Eq.~\eqref{eq_QG_density}), which involves `tails' of $U(1)$ densities. We leave a detailed study of the fluctuating hydrodynamics of systems with QG symmetry for future work. One further challenge for this theory will be to capture the unusual finite size effects discussed in Sec.~\ref{sec_finite_size}.

%======================================================================%
%======================================================================%
%======================================================================%
\subsection{Realization in classical spin chains}
Quantum group symmetry plays a role in classical systems as well. To show this, we construct a classical lattice model with $U_q(sl_2)$ symmetry.
We start from the  classical Heisenberg chain, which is $SU(2)$ symmetric
\begin{equation} \label{eq:classical_Heis}
    H=  \frac12\sum_i \vec S_i \cdot \vec S_{i+1},
    \qquad
    |\vec S_i|=1 ,
\end{equation}
where the spins obey the Poisson brackets
\begin{equation}
    \{S^a_i,S^b_j\}=\epsilon^{abc} \delta_{ij}S^c_i ,
\end{equation}
and time evolution is generated by
\begin{equation}
    \frac{d}{dt}S^a_i = \{ S^a_i,H \}\, .
\end{equation}
This model is $SU(2)$ symmetric: $\{\sum_i S^\pm_i, H\}=\{\sum_i S^z_i, H\} = 0$, with $S^\pm = S^x \pm i S^y$. The thermalization of this model has been debated for decades: it is understood to exhibit conventional diffusion, albeit with sizeable finite time corrections \cite{Glorioso:2020loc,McRoberts:2021jmb}.

We now deform this model to construct a classical model which has $U_q(sl_2)$ QG symmetry. We define
\begin{equation} \label{eq:classical_generators}
    K_j=e^{\eta S^z_j},
    \ \ 
    E_j= f(S^z_j) S^+_j,
    \ \ 
    F_j=\frac{-\eta}{\sinh\eta}\, f(S^z_j) S^-_j,
\end{equation}
where the explicit form of the function $f$ can be found in App.~\ref{app:classical}. These generators then satisfy the quantum group Poisson brackets with parameter $q=e^{\eta}$. Using these as building blocks, we can construct the Hamiltonian
\begin{align}\label{eq_classical_QG_model}
H = - &\sum_{i=1}^{L-1}\bigg[ %]
K_i^{-1/2}
\left(
E_{i+1}F_i
+
E_iF_{i+1}
\right)
K_{i+1}^{1/2}
\\
&-
\frac{1}{\eta \sinh \eta} K_i^{-1}K_{i+1}
+ \frac{1}{\eta \tanh \eta} (K_i^{-1}+K_{i+1})\notag
%\left(K_i^{-1}K_{i+1} -\cosh\eta\, K_{i+1} -\cosh\eta\, K_i^{-1}\right) 
\bigg]
\end{align}
which is quantum group symmetric. As $\eta \to 0$, the Hamiltonian reduces to the Heisenberg model up to a constant shift, and can be thought as a deformation of it which preserves quantum group symmetry.

\begin{figure}[t]
\includegraphics[width=0.85\columnwidth]{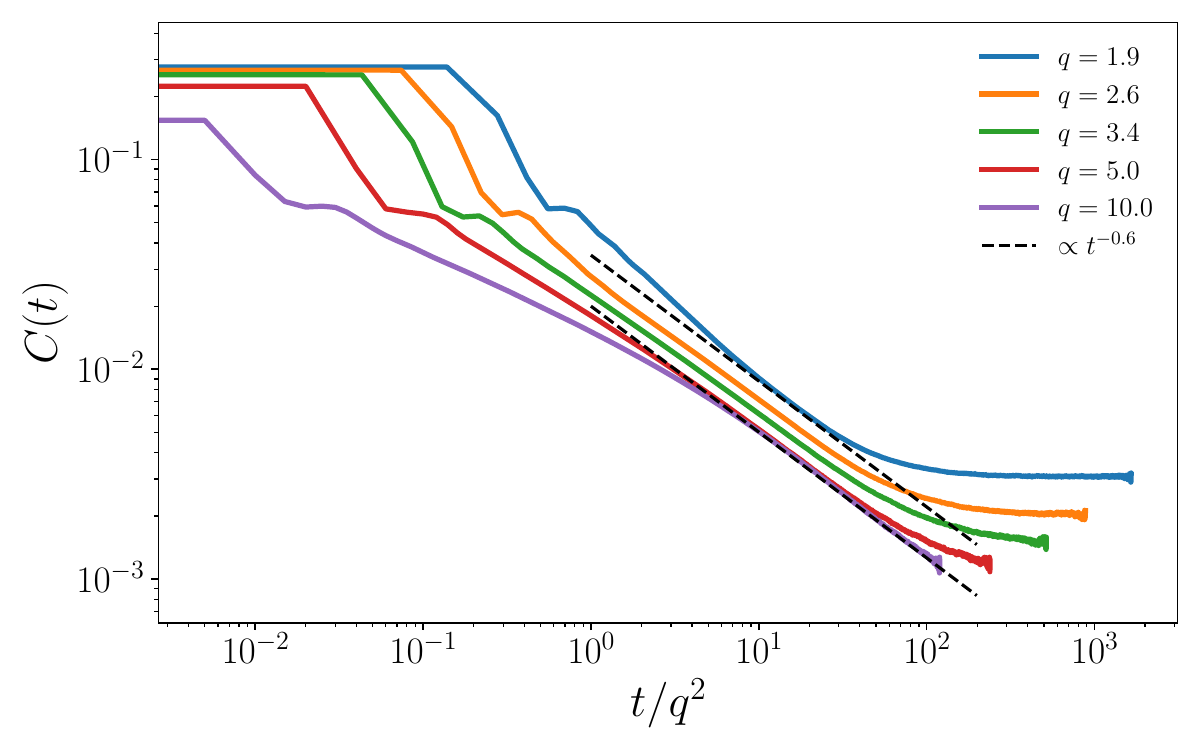}
\caption{Infinite temperature autocorrelation function for transverse spin in QG symmetric classical models with various $q$,
for system size $L = 50$.  Dashed line indicates the scaling $1/t^{3/5}$ for guide to the eye.}
\label{fig_C_XXt_Cl}
\end{figure}

We study the model \eqref{eq_classical_QG_model} numerically using a fourth order Runge-Kutta method to time evolve the system. Correlation functions are computed by starting from a random configuration of spins, and then averaged over many different initial configurations. The results, shown in Fig.~\ref{fig_C_XXt_Cl}, confirm power-law decay of the transverse spin in models with QG symmetry. 

One advantage of this classical model is that it is expected to remain non-integrable for any value of $q$, unlike the deformed quantum XXZ model \eqref{eq_HQG} which becomes integrable when $q\to \infty$. This allows to better test for the possible $q$-dependence of exponents. Fig.~\ref{fig_C_XXt_Cl} shows that the power-law exponent in the autocorrelation function is essentially unchanged while $q$ varies over an order of magnitude, with an exponent consistent with the one observed in the quantum models. This suggests that generic $U_q(sl_2)$-symmetric spin chains are all described by the same superdiffusive universality class as long as $q\neq 1$.

%######################################################################%
%======================================================================%
%======================================================================%
%======================================================================%
%######################################################################%
\section{Finite size scaling and thermodynamics}\label{sec_finite_size}

We found that QG symmetry leads to power-law decay of local operators such as transverse spin, with an exponent that may differ from known dissipative universality classes arising in models with conventional symmetries. These observations were made in regimes with negligible finite size effects. However, we will see that the most striking difference between observables in spin chains with conventional versus QG symmetry is in the very late time saturation of correlation functions. We start by reviewing finite size effects in systems with conventional hydrodynamics before turning to the unusual behavior of QG symmetric spin chains.

Finite size effects are usually not the focus of numerical studies in chaotic quantum many-body systems, and are often only used as consistency checks that observables of interest have reached their thermodynamic limit (exceptions in translation-invariant systems include \cite{Friedman:2019gyi,Winer:2020gdp, PhysRevB.104.054415,Maceira:2024syz}). However, finite size effects are a valuable resource in the study of hydrodynamic transport. Indeed, they should be entirely fixed by the hydrodynamic theory and its boundary conditions. For example, diffusion on a 1d system of size $L$ with periodic boundary conditions leads to the finite size autocorrelation function (ignoring intermediate time power-law corrections):
\begin{equation}\label{eq_Fscaling_finitesize}
\langle n(t,x=0)n\rangle
	\simeq \frac{1}{L} \sum_{q\in \frac{2\pi}{L}\mathbb Z}\chi e^{-D q^2 |t|}
	= \frac{\chi}{L} F(\tau)\, ,
\end{equation}
where $\tau\equiv D t \left(\frac{2\pi}{L}\right)^2$ and the scaling function $F(\tau) \equiv \sum_{n\in \mathbb Z} e^{-n^2 |\tau|}$ connects conventional hydrodynamic power-law decay $F(\tau) \simeq \sqrt{\pi/\tau}$ when $\tau \ll 1$ to late time saturation $F(\infty) = 1$. $F$ is universal: it only depends on the universality class (here, diffusion) and the boundary conditions (here, periodic). We emphasize that the cross-over from $\langle n(t)n\rangle\sim 1/\sqrt{t}$ to $\langle n(t)n\rangle\sim 1/L$ at late times $t\gg L^2$ is a {\em prediction} of diffusion, rather than signaling its breakdown. In particular, even when the exact scaling function $F$ is not known, searching for collapse of finite size data in terms of $t/L^z$
\begin{equation}\label{eq_Fscaling_finitesize_gen z}
\langle n(t,x=0)n\rangle
	= \frac{1}{L^d} F(t/L^z)\, ,
\end{equation}
with $d$ the spatial dimension, is a much more cost-efficient way to identify the exponent $z$ of anomalous hydrodynamics compared to using the full position resolved data \eqref{eq_Ctx}. We confirm in App.~\ref{app_finitesize} finite size data collapse in diffusive models with periodic boundary conditions agrees with the universal scaling function $F(\tau)$ from Eq.~\eqref{eq_Fscaling_finitesize}.

\begin{figure}[t]
\includegraphics[width=0.85\columnwidth]{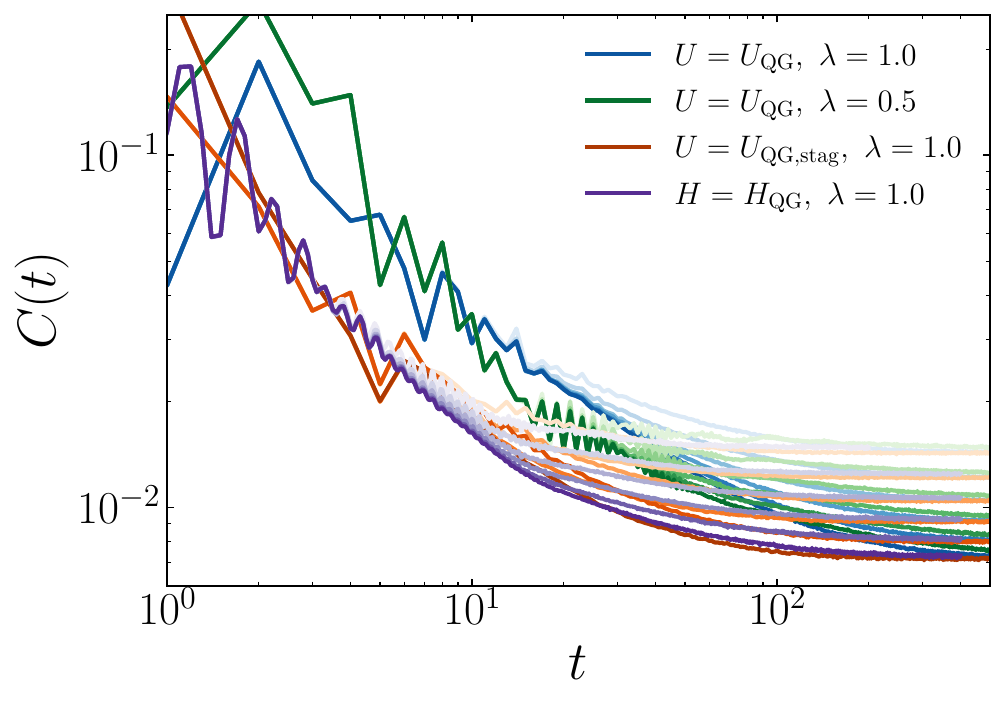}
\caption{Infinite-temperature autocorrelation functions $C(t)=\mathrm{Tr}(\mathcal{O}(t)\mathcal{O})$ for transverse spin $\mathcal{O}=\sigma^x$ at fixed $q=2.6$ and different values of $\lambda$, for both the Floquet and Hamiltonian systems, and for system sizes $L=20,22,24,26,28,30$ (colors from light to dark).}
\label{fig_diff_lambda}
\end{figure}

Let us now turn to spin chains with QG symmetry. Fig.~\ref{fig_diff_lambda} shows clear finite size effects at late times in quantum Hamiltonian and Floquet models with QG symmetry. Moreover, different models with the same value of the QG symmetry parameter $q$ and size $L$ appear to converge to the same infinite time value of the correlator
\begin{equation}
C(\infty) \equiv \lim_{T\to \infty}\frac1{T}\int_0^T dt  \, \langle \sigma^+_{i}(t)\sigma^-_{i}(0)\rangle\, .
\end{equation}
Thus $C(\infty)$ only depends on $q$ and $L$, and is independent of the Hamiltonian (or Floquet operator) at infinite temperature $\beta=0$. This parallels what happens in conventional diffusive systems, where the charge susceptibility $\chi$ appearing in \eqref{eq_Fscaling_finitesize} involves infinite temperature traces of densities and is hence Hamiltonian independent.

\begin{figure}[t]
\includegraphics[width=1.0\columnwidth]{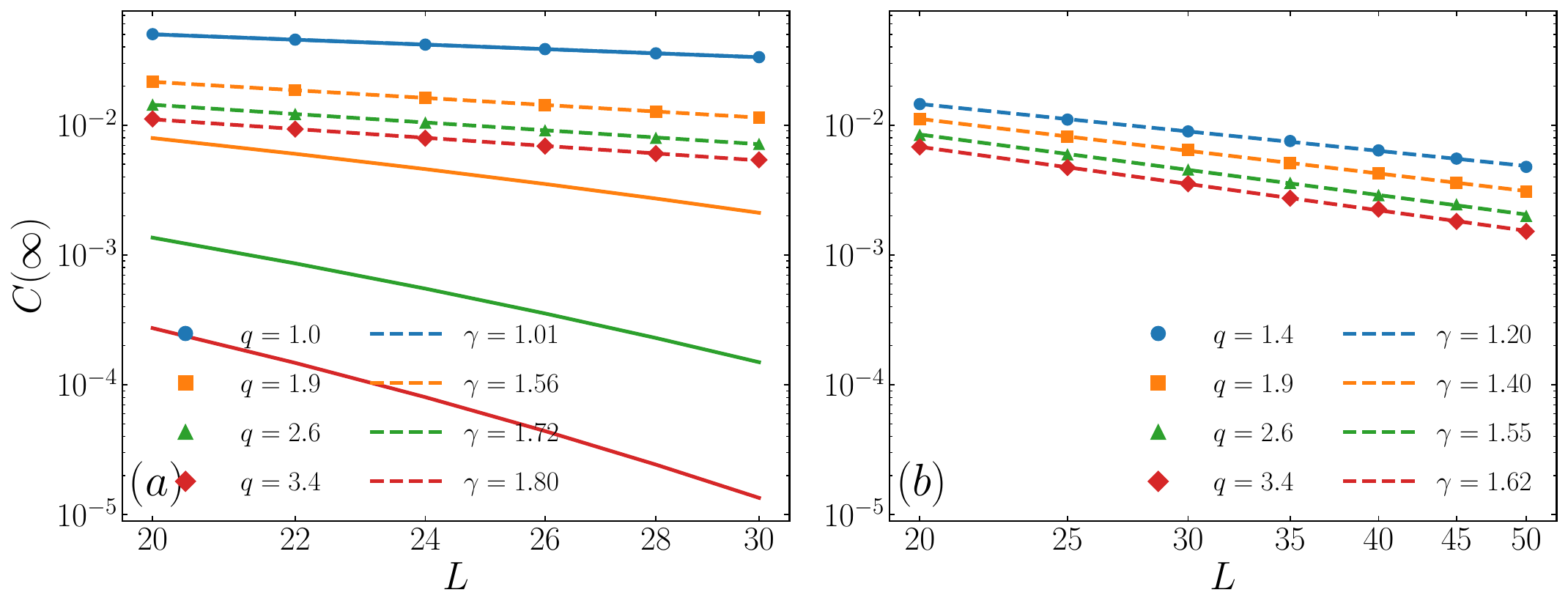}
\caption{Scaling of the long-time limit $C(\infty)$ with system size $L$ for various values of $q$. 
Results are shown for 
(a) the quantum Floquet model at $\lambda = 1.0$ and 
(b) the classical model.
Dashed lines indicate fits of the form $\sim L^\gamma$.
Solid lines indicate the Mazur bound.}
\label{fig_finite_size}
\end{figure}

However, in stark contrast to systems exhibiting conventional fluctuating hydrodynamics or diffusion, the late time value $C(\infty)$ does not scale as $1/L$ with system size, as shown in Fig.~\ref{fig_finite_size}. Thus the autocorrelation function will not collapse to the form \eqref{eq_Fscaling_finitesize_gen z}, for {\em any} value of $z$, unlike models with diffusive, superdiffusive, or subdiffusive hydrodynamic modes protected by conventional continuous symmetries. The behavior observed in Figs.~\ref{fig_finite_size} is also markedly different from what is expected for correlators of non-conserved operators, whose late time saturation value is expected to be inversely proportional to the Hilbert space dimension $C(\infty) \sim 1/\mathcal D = 2^{-L}$ (through the same argument that produces the late time value of the spectral from factor \eqref{fig_SFF})---the contrast is particularly clear in Fig.~\ref{fig_money}.

A simple analytical argument suggests that $C(\infty)$ for quantum group charges should indeed interpolate between the conventional $1/L$ behavior for conserved densities and $2^{-L}$ behavior for non-conserved operators. $C(\infty)$ is lower bounded by overlaps with conserved quantities (Mazur bound). For an operator $\mathcal{O}$ that overlaps with a conserved charge $Q$ ($[H,Q] = 0$), one has
\begin{equation} \label{eq:Mazur1}
C(\infty) = \lim_{T\to \infty}\frac{1}{T}\int_0^T dt \, \langle \mathcal{O}^\dagger(t) \mathcal{O}\rangle 
	\geq \frac{\langle \mathcal{O}^\dagger Q\rangle\langle Q^\dagger\mathcal{O} \rangle}{\langle Q^\dagger Q\rangle}
\end{equation}
with $\langle \cdot \rangle = \Tr(\rho \ \cdot \ )$. This inequality can be generalized (and strengthened) by considering additional charges, see App.~\ref{app_Mazur}. Such bounds are often used in integrable models to show the existence of a non-zero Drude weight, in which case the operator $\mathcal{O}$ is taken to be the current. In a conventional diffusive system, applying it to the conserved density $\mathcal{O} = n_i$ and taking $Q = \sum_i n_i$, one finds $C(\infty) \geq \langle Q Q\rangle/L^{2d} = \chi /L^{d}$, which is saturated by the diffusive behavior \eqref{eq_Fscaling_finitesize}. Let us now apply it to the transverse spin $\mathcal{O} = \sigma_i^-$ in a model with QG symmetry. A natural candidate for a conserved charge is the QG charge $Q = E = \sum_i \hat \sigma^+_i$, defined by
\begin{equation}\label{eq_QG_density}
\hat \sigma^+_i = \cdots  \, 
	q^{-\frac12 \sigma^z_{i-2}}   \, 
	q^{-\frac12 \sigma^z_{i-1}}  \,
	\sigma^+_i \,
	q^{\frac12 \sigma^z_{i+1}}  \, 
	q^{\frac12 \sigma^z_{i+2}}  \,
	\cdots\, .
\end{equation}
This `density' reduces to the $SU(2)$ density $\hat \sigma^+_i = \sigma_i^+$ when $q=1$, but is dressed by a topological tail in general. It commutes with the Hamiltonians we study (in fact there are several choices of such densities, see App.~\ref{app_Mazur}). This charge overlaps with the local transverse spin, leading to the bound (at infinite temperature, $\rho = \mathds 1 / 2^{L}$)
\begin{equation} \label{eq:Mazur2}
C(\infty) \geq \frac{\langle \sigma^-_i E\rangle \langle E^\dagger \sigma^+_i\rangle}{\langle E^\dagger E\rangle}
	= \frac{1}{L} \left( \frac{q + q^{-1} +2}{2(q+q^{-1})}\right)^{L-1}.
\end{equation}
Interestingly the right-hand side interpolates between $1/L$ when $q=1$, in which case it is the optimal bound for conventional $SU(2)$ diffusion, and $\sim 2^{-L}$ for $q=\infty$. This bound is compared to the numerical results in Fig.~\ref{fig_finite_size}(a)---while it is far from being tight for $q>1$, it establishes that the late time floor of the transverse spin autocorrelation function must be parametrically larger than the value $2^{-L}$ expected for a non-conserved operator. We discuss in App.~\ref{app_Mazur} how to strengthen the bound by considering additional charges built out of $E_i$ and $S_z$.

It is interesting to contrast the result \eqref{eq:Mazur2} in QG symmetric models with models exhibiting Hilbert space fragmentation, where Mazur bounds show that the late time saturation of local correlators $C(\infty) \sim 1/\sqrt{L}$ is instead parameterically larger than for conserved densities in generic chaotic models \cite{Rakovszky:2020apb,Moudgalya:2021ixk,Lehmann:2022gds,Hart:2023zgl}.

%~ \LD{is the late time value approached exponentially, like in diffusion? $\sim \frac{1}{L} \left(1+e^{-t/L^2} + ...\right)$}

%######################################################################%
%======================================================================%
%======================================================================%
%======================================================================%
%######################################################################%
\section{Implications for integrable models}\label{sec_integrable}

While the main emphasis of this work is that QG symmetry has important consequences for the local dynamics of generic chaotic many-body systems, some of these consequences may extend to integrable models as well. We focus on the integrable XXZ model in this section and find that this is the case. These results show the importance of extending the generalized hydrodynamic description of integrable models (GHD) \cite{Castro-Alvaredo:2016cdj,Bertini:2016tmj,Doyon:2023zcl} to incorporate QG symmetry.

%======================================================================%
%======================================================================%
%======================================================================%
\subsection{(Super)diffusion of transverse spin in $\Delta > 1$ XXZ}

\begin{figure}[t]
\includegraphics[width=1\columnwidth]{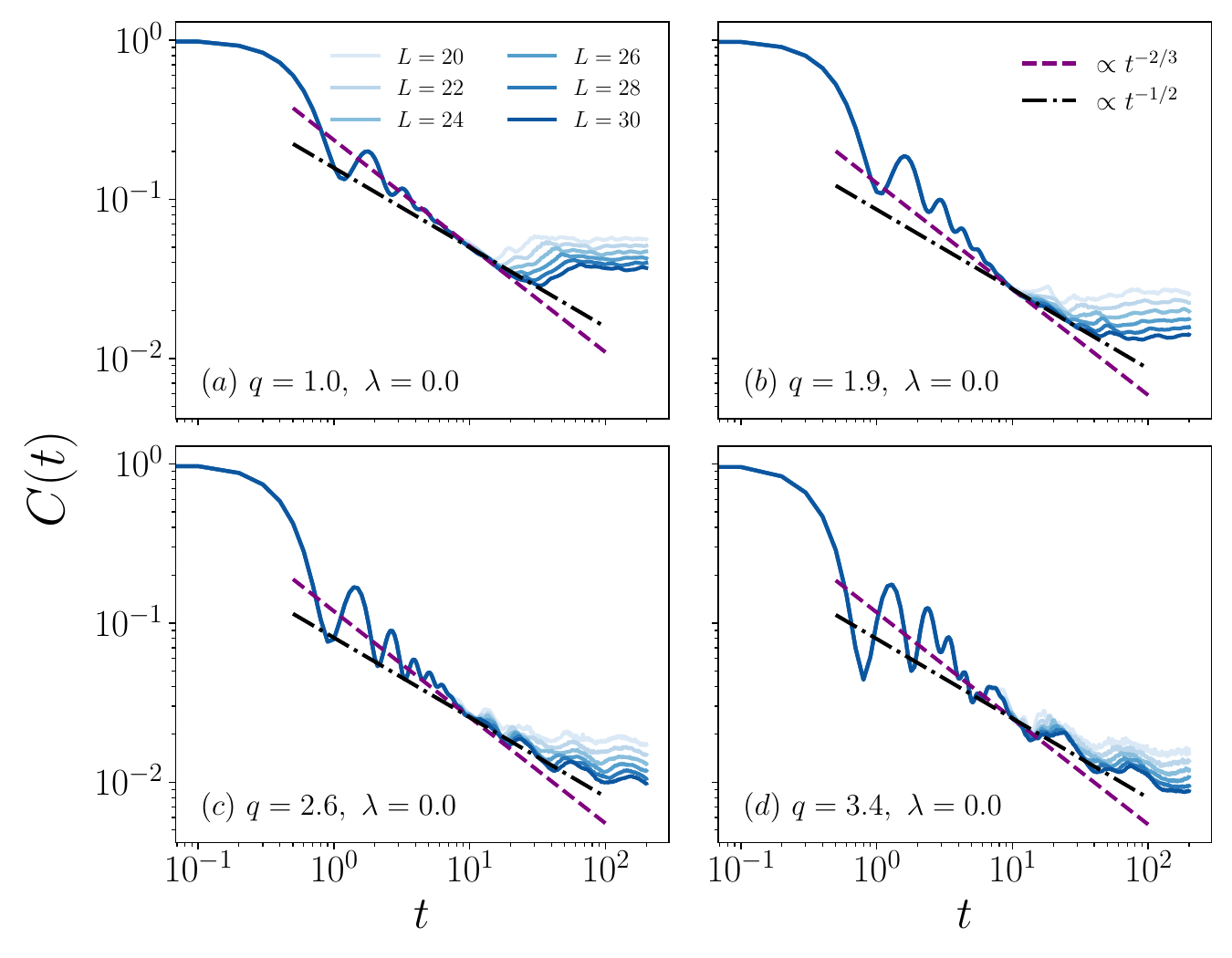}
\caption{Infinite temperature autocorrelation function $C(t)= \Tr(\mathcal O(t)\mathcal O)$ for transverse spin $\mathcal O = \sigma^x$ in the integrable XXZ model, with $\Delta = \frac12(q+q^{-1})\geq 1$,
for system size $L = 20,22,24,26,28,30$ (color from light to dark). Dotted ($1/t^{1/2}$) and dashed ($1/t^{2/3}$) lines are guides to the eye.}
\label{fig_C_XXt_Int}
\end{figure}

As discussed in Sec.~\ref{sse_autocorrelation}, transverse spin dynamics in the integrable XXZ model has received little attention. We find that, like in chaotic models with QG symmetry, it features surprising long-lived behavior even though it does not overlap with a conventional conserved density. We start by considering the easy axis-regime $\Delta > 1$. Fig.~\ref{fig_C_XXt_Int} shows that the transverse spin autocorrelation function again exhibits slow polynomial decay. The exponent $z$ appears to be superdiffusive, possibly in the same universality class as the one observed in chaotic models with QG symmetry. However, ruling out diffusion with large intermediate time corrections---as is done in App.~\ref{app_ruleoutdiff} for chaotic models---is likely more tricky in integrable models where corrections can arise from many modes, and will require further investigation. We have observed similar behavior in a classical discrete-time version of the XXZ model \cite{Krajnik:2021anis}.

One notable difference between the autocorrelation function shown in Fig.~\ref{fig_C_XXt_Int} and the corresponding observable in chaotic systems (Figs.~\ref{fig_C_XXt_Hamiltonian} and \ref{fig_C_XXt_Floquet}) is its finite size effects, which are more featured in the integrable model. This suggests an intricate coupling between the QG density and some of the many other conserved densities in the integrable XXZ chain.

%======================================================================%
%======================================================================%
%======================================================================%
\subsection{Evidence for transverse spin sound modes in $\Delta < 1$ XXZ}

The chaotic systems we have studied so far always had $U_q(sl_2)$ symmetry with $q\in \mathbb R$: indeed, $q\notin \mathbb R$ would have led to non-unitary models. Instead, the XXZ model in the easy-plane regime $\Delta < 1$ realizes the QG symmetry with complex parameter on the unit circle $q = e^{i\theta}$, $\theta\in (0,2\pi)$, while still being unitary in the bulk. Indeed, Eq.~\eqref{eq:h0} is still Hermitian in this case, except for the boundary term. We choose to preserve exact unitarity of the model in the finite size numerics by dropping the boundary term, and hence break QG symmetry at the boundary.

\begin{figure}[t]
\includegraphics[width=1\columnwidth]{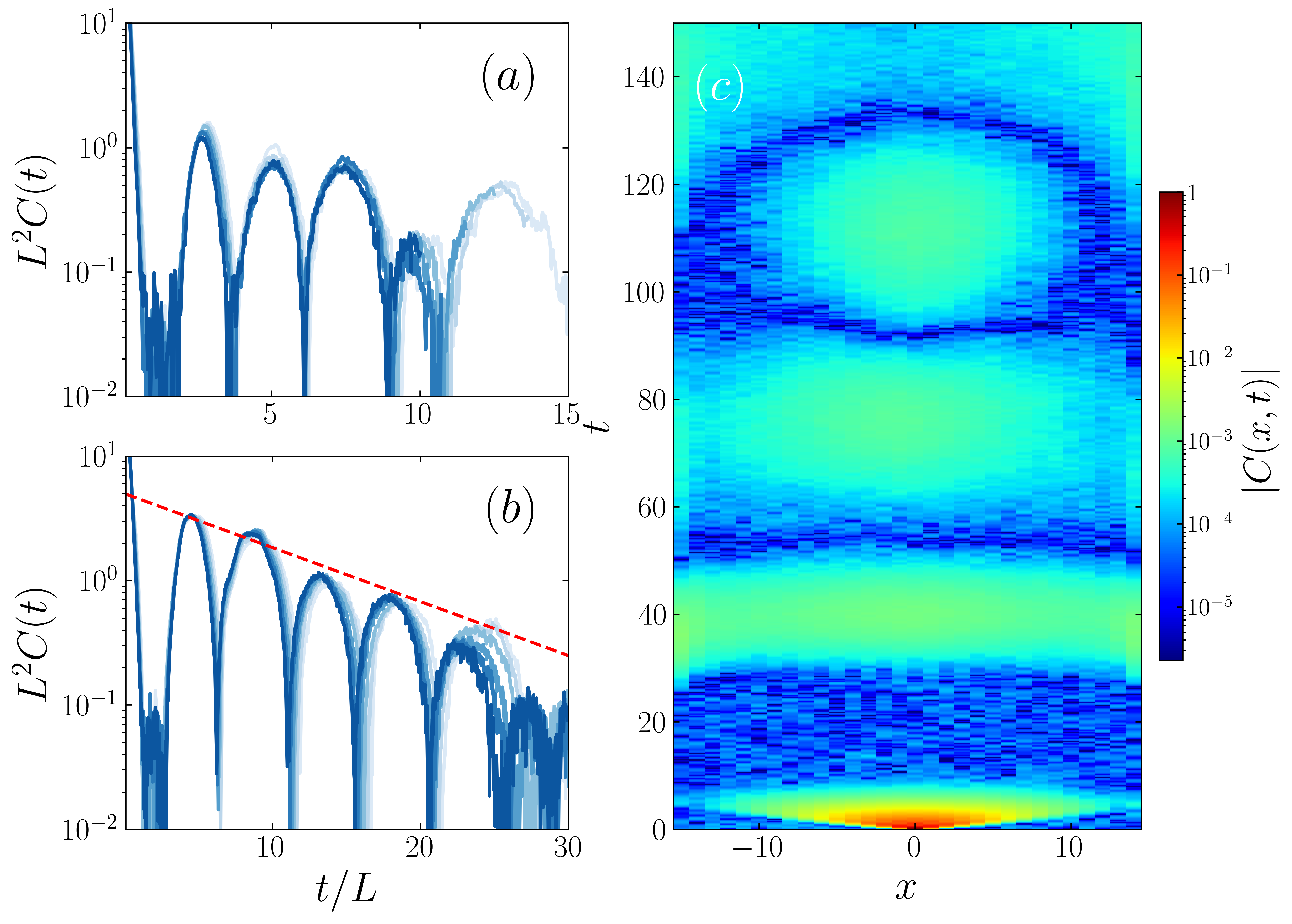}
\caption{Infinite temperature autocorrelation function $C(t)= \Tr(\mathcal O(t)\mathcal O)$ for transverse spin $\mathcal O = \sigma^x$ in the integrable XXZ model, with $\Delta = \frac12(q+q^{-1}) = 0.6$,
for system sizes $L = 20,22,24,26,28,30$ (from light to dark)
for the Hamiltonian (a) and Floquet model (b). While the early time behavior is a fast exponential, finite size revivals appear at very late times $t\sim L$. The dashed line $\exp(-0.1t/L)$ emphasizes the slow exponential decay of these peaks.
(c) Heat map of $C(t,x)$ in the $(x,t)$ plane for Hamiltonian model with $L = 30$. }
\label{fig_int_easyplane}
\end{figure} 

Previous numerical studies have found exponentially decaying behavior of the transverse spin \cite{PhysRevB.49.15669}, which we confirm, see Fig.~\ref{fig_int_easyplane}(a,b). However, intriguing finite size effects suggest the existence of a ballistic or sound-like mode: notice the late time `revivals' appearing in the autocorrelation function at regular time intervals $n L / v$, with $n=1,2,3,\ldots$ and $v\approx 2$. The linear scaling of these revivals with system size $L$ suggests a ballistic or sound-like mode propagating with velocity $v$ and bouncing off the edges. Moreover, the extremely slow exponential decay is consistent with the symmetry only being broken at the boundary, where the conserved charge can leak out.

The spatially resolved correlation function entirely challenges this interpretation: Fig.~\ref{fig_int_easyplane}(c) shows no sign of a linearly dispersing mode, and the revival is in fact smeared throughout the entire system! One possibility is that this is due to the local transverse spin $\sigma^+_i$ imperfectly overlapping with the (likely nonlocal) density propagating ballistically. Recall that the QG density \eqref{eq_QG_density} lives at the end of a topological string. The absence of a (super)diffusive mode for the $\Delta < 1$ XXZ model suggests that this density couples with additional conserved charges, which may also be nonlocal. It is possible that the ballistic mode can be resolved by a two point function $\langle \hat \sigma^+_j \sigma^-_i\rangle$, where $\hat \sigma^+_j$ carries unit $U(1)$ charge like $\sigma^+_j$, but is additionally attached to a string connecting it to the other operator at $i$. We leave testing this hypothesis and further exploring the consequences of QG symmetry in integrable models for future work.

%######################################################################%
%======================================================================%
%======================================================================%
%======================================================================%
%######################################################################%
\section{Discussion}\label{sec_discuss}

We have found that quantum group symmetry leads to distinctive hydrodynamic signatures in both chaotic and integrable systems, quantum or classical, with either Hamiltonian or Floquet dynamics. This opens several interesting paths for future research.

First, our finding motivates extending effective field theories for fluctuating hydrodynamics (as well as GHD) to accommodate quantum group symmetry. Other generalized continuous symmetries can be seamlessly integrated in hydrodynamic EFTs, see, e.g., Refs.~\cite{Grozdanov:2016tdf,Armas:2018ibg,Glorioso:2018kcp,Delacretaz:2019brr,Iqbal:2020lrt,Delacretaz:2026owo}. We expect the generalization to quantum group symmetry to be more challenging---given the absence of a local current, and our observation of unusual finite size effects---but also more rewarding, since it appears to allow for large hydrodynamic fluctuations and a possibly new superdiffusive universality class.

Furthermore, this new mechanism for superdiffusion involving exotic symmetries motivates revisiting models showing anomalous hydrodynamics with exponent $z$ that is not explained by conventional symmetries. We have in mind in particular the anomalous subdiffusion $2<z<4$ observed in Fredkin and Motzkin chains \cite{Richter:2021oxe,Singh:2021tfm,McCarthy_2025}; the possible relevance of nonlocal conserved densities \cite{McCarthy_2025} and Temperley-Lieb algebra \cite{Salberger:2018ltz} in these models suggests connections with our results. In a similar vein, it would be interesting to establish superdiffusion of heat in the PXP model \cite{Ljubotina:2022ssg} (as opposed to conventional diffusion with power-law corrections) using the simple diagnostic presented in Eq.~\eqref{eq_R_ratio_gen}. It would also be interesting to explore other models with QG symmetry \cite{Batchelor:1989uk,deVega:1993ae,Quella:2020nmv,Franke:2024vdo}.

Our results show that quantum group symmetry has important implications on local observables in non-integrable models, suggesting it may be a promising generalization of continuous global symmetry in many-body systems, as advocated in Refs.~\cite{Gabai:2024puk,Gabai:2024qum, Gorbenko:2025wzs}. However, many questions remain open on the scope of this generalization: can QG symmetry emerge from microscopics that does not have the symmetry? How natural is it? Can it be spontaneously broken, and would a Goldstone-like theorem apply? Can these symmetries be gauged? The answer to the last question is positive, at least for particular values of $q$ where the fusion algebra of representations truncates, in which case quantum groups can be used to regulate lattice gauge theories in a Hamiltonian framework \cite{Zache:2023dko}. Such models can be thought of lattice gauge theories based on a quantum group as a gauge group, also for generic $q$. It would also be interesting to see if continuous $q$-deformed gauge theories exist beyond two dimensions.

Finally, XXZ-like models such as the ones we studied have been realized in landmark cold atom experiments \cite{Jepsen:2020cjl,Wei:2021wko,Scholl:2021xeo}, see in particular Ref.~\cite{PhysRevX.11.041054} which probed transverse spin response for $\Delta< 1$. Our surprising predictions for transverse spin dynamics both in the integrable XXZ model at $\Delta > 1$ and some of its chaotic deformations motivate further investigating these observables in experiments.

%######################################################################%
%======================================================================%
%======================================================================%
%======================================================================%
%######################################################################%
\section*{Acknowledgements}

We thank Sihan Chen, Barak Gabai, Sarang Gopalakrishnan, Nick Holfester, Ruchira Mishra, and Jiaxin Qiao for helpful discussions. LVD is supported by a NSF CAREER award (DMR-2441227) and a Sloan Fellowship. We also thank the organizers of the workshop ``Effective Field Theories for Hydrodynamics'' at the Bernoulli Center, during which the project was initiated.

\bigskip

%######################################################################%
%======================================================================%
%======================================================================%
%======================================================================%
%######################################################################%
\section*{Appendix}
\appendix

% Make appendix subsections look like subsubsections in the main text
\makeatletter
% Copy REVTeX's definition of \subsubsection and apply it to \subsection
\renewcommand\subsection{\@startsection{subsection}{2}{0pt}%
  {1.0ex plus .5ex minus .2ex}%
  {0.5ex plus .2ex}%
  {\normalfont\normalsize\itshape\centering\medskip}}
\makeatother

%######################################################################%
%======================================================================%
%======================================================================%
%======================================================================%
%######################################################################%
\section{Ruling out conventional diffusion}\label{app_ruleoutdiff}

Transverse spin appears to decay superdiffusively in spin chains with QG symmetry: $\langle \sigma^+_i(t) \sigma^-_i\rangle \sim 1/t^{d/z}$ with $d=1$ and $z\approx 1.6 <2$. However, it can be challenging in intermediate time numerics to distinguish between superdiffusion and vanilla diffusion with power law corrections, which are always present and only have a relative $1/\sqrt{t}$ suppression in $d=1$ spatial dimension. In this appendix, we explain how this possibility can be ruled out using detailed predictions from the effective field theory (EFT) of diffusion (see \cite{Delacretaz:2026owo} for a review).

The EFT of diffusion has two cubic vertices, related to derivatives of transport parameters with respect to the associated densities: $D' \equiv dD/dn$, $\chi' \equiv d\chi/dn$. The former is responsible for the leading correction to two-point functions through a 1-loop diagram:
\begin{equation}\label{eq_nn}
\langle n(t)n\rangle
	= \frac{\chi }{(4\pi D t)^{1/2}} \left[1 + \frac{\chi D'^2}{4\sqrt{\pi} D^{5/2}} \frac{1}{\sqrt{t}} + O(\tfrac{1}{t})\right].
\end{equation}
Another useful prediction of the EFT is that the same coupling responsible for this correction ($D'$) also enters in the leading diffusive connected 3pt function
\begin{equation}\label{eq_nnn}
\langle n(2t) n(t) n\rangle_c
	= \frac{\chi ^2}{4\pi Dt} \left( \frac12\frac{\chi'}{\chi} + \frac{D'}{D} (1+\sqrt{2})\right) + O(\tfrac1{t^{3/2}})\, .
\end{equation}
The two couplings $\chi'$ and $D'$ can be furthermore extracted independently by considering the spatially resolved 3pt function, see Ref.~\cite{Delacretaz:2023ypv} for details. Having extracted $D'$ from the 3pt function, it is then possible to find to find the size of the correction in \eqref{eq_nn}
\begin{equation}
\frac{\sqrt{\tau_{\hbox{\scriptsize 1-loop}}}}{\sqrt t} \equiv\frac{\chi D'^2}{4\sqrt{\pi} D^{5/2}} \frac{1}{\sqrt{t}}\, ,
\end{equation}
for the time scales $t$ accessible numerically.

\begin{figure}[t]
\includegraphics[width=0.85\columnwidth]{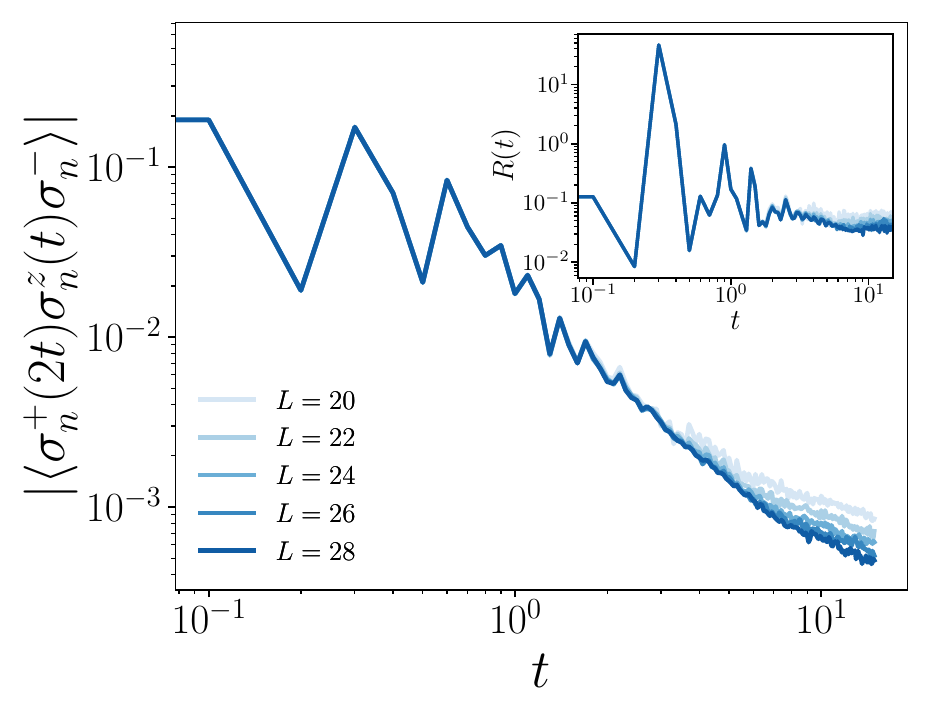}
\caption{Infinite temperature 3pt correlation function $\sigma_{n}^{+}(2t)\sigma_{n}^{z}(t)\sigma_{n}^{-}$ in the QG symmetric Hamiltonian models and $R(t)$ (defined in \eqref{eq_R_ratio}) in the inset, for $q = 2.6$, $\lambda = 1.0$ and system sizes $L = 20,22,24,26,28$.}
\label{fig_a1}
\end{figure}

This is a quantitative prediction for the size of corrections in models with a single diffusing charge, but the same idea provides a qualitative test that is applicable much more generally, for any number of diffusing charges in any dimension. Since the leading 1-loop correction always arises from cubic vertices that can produce connected 3pt functions, the dimensionless ratio of 3pt and 2pt functions estimates the size of these corrections
\begin{equation}\label{eq_R_ratio_gen}
\left(\frac{\tau_{\hbox{\scriptsize 1-loop}}}{t}\right)^{d/2}  \sim R(t) \equiv \frac{\langle n(2t)n(t)n\rangle_c^2}{\langle n(t)n\rangle^3}\, .
\end{equation}
The densities on the right-hand size could also be chosen to be separated in space to remove $\chi'$-like contributions in \eqref{eq_nnn}, but we will not do so. Let us apply this to the models with QG symmetry. The only 3pt function allowed by $U(1)$ symmetry is $\langle \sigma^+ \sigma^- \sigma^z\rangle$. We consider the following ratio
\begin{equation}\label{eq_R_ratio}
R(t)
	\equiv \frac{\langle \sigma^+(2t) \sigma^z(t) \sigma^-(0)\rangle^2}{\langle \sigma^+(t)\sigma^-(0)\rangle^2 \langle \sigma^z(t)\sigma^z(0)\rangle}\, .
\end{equation}
Fig.~\ref{fig_a1}  shows that this ratio is $R =$ 0.04-0.05 at times $t=$ 3-6 before finite size effects start in Fig.~\ref{fig_C_XXt_Hamiltonian}. This would thus lead to a 4-5$\%$ error in the determination of $z\simeq 2 - 2R$ in this time window, if the system were diffusive. This can thus not account for the 20\% deviation from $z=2$ observed across models, and rules out loop corrections to diffusion as the source of apparent superdiffusion in QG symmetric models.

%######################################################################%
%======================================================================%
%======================================================================%
%======================================================================%
%######################################################################%
\section{Further data on dynamics in QG symmetric spin chains}\label{app_data}

\begin{figure}[t]
\includegraphics[width=1.0\columnwidth]{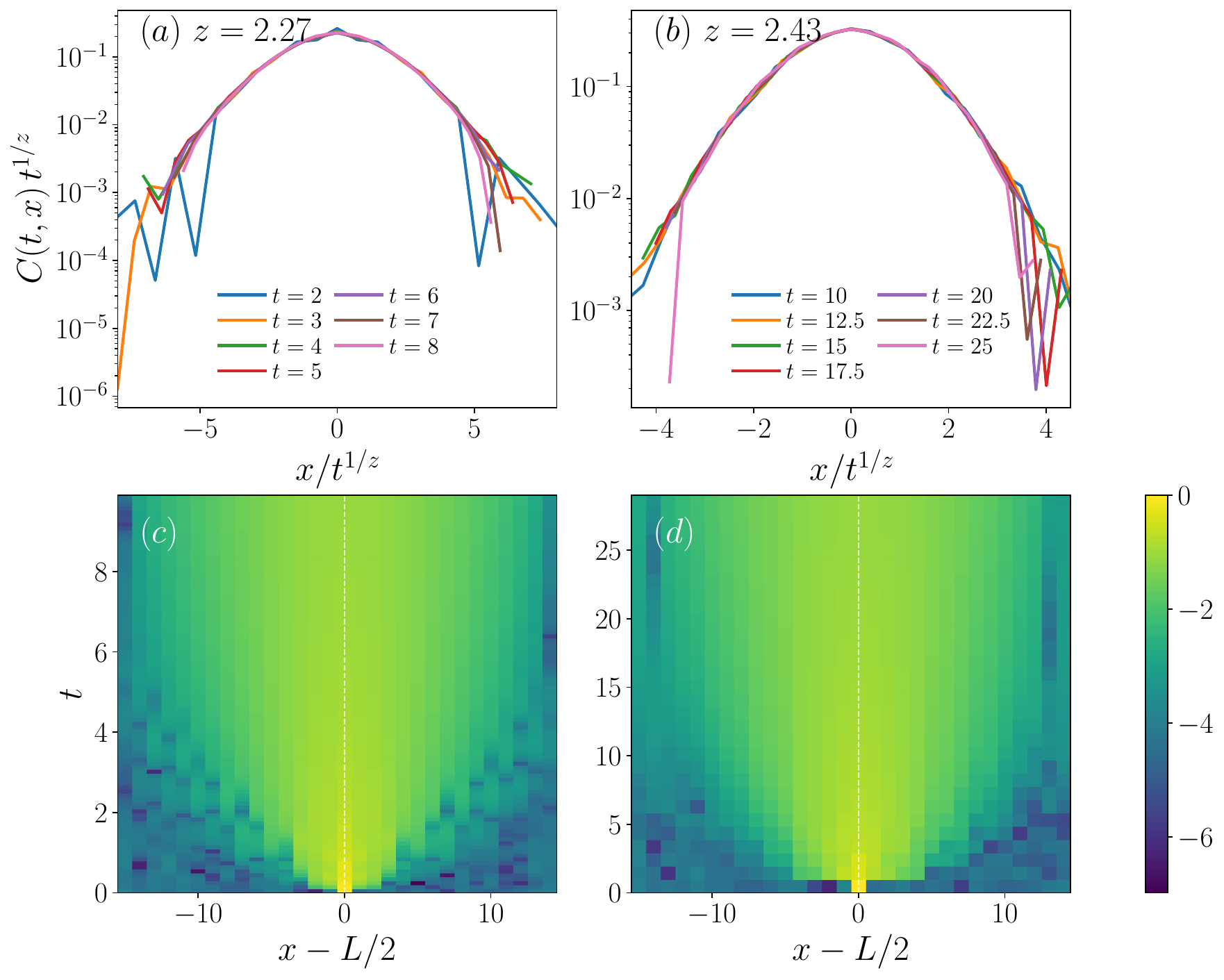}
\caption{Scaling collapse of $C(t,x)t^{1/z}$ versus $x/t^{1/z}$ for
longitudinal spin ${\cal O} = \sigma^z$ in
the (a) Hamiltonian model and (b) Floquet model for $q = 2.6$ and $\lambda = 1.0$. The corresponding heat maps of $C(t,x)$ in the $(x,t)$ plane are shown in (c) and (d), respectively. The system size is $L = 30$.}
\label{fig_b1}
\end{figure}

In this appendix, we consider additional observables beyond the transverse spin autocorrelation function to help guide future fluctuating hydrodynamic theory constructions incorporating QG symmetry. We start with the position-resolved longitudinal spin correlation function $C_{zz}(t,x=j-i) \equiv \langle\sigma^z_j(t)\sigma^z_i\rangle$. Fig.~\ref{fig_b1} shows data collapse of the scaling function. Interestingly, it appears to be slightly subdiffusive. We believe that this may be due to intermediate time effects: indeed, the finite-size collapse observed in Fig.~\ref{fig_e1}---which is less sensitive to intermediate time corrections---shows very good agreement with diffusive scaling $z=2$. Nevertheless, our current numerics alone cannot rule out slightly subdiffusive dynamics of longitudinal spin.

%~ This suggests that the relevant hydrodynamic nonlinearities that are likely present in the transverse spin sector only give irrelevant corrections to longitudinal spin dynamics. \LD{further comment fig?}

\begin{figure}[t]
\includegraphics[width=1.0\columnwidth]{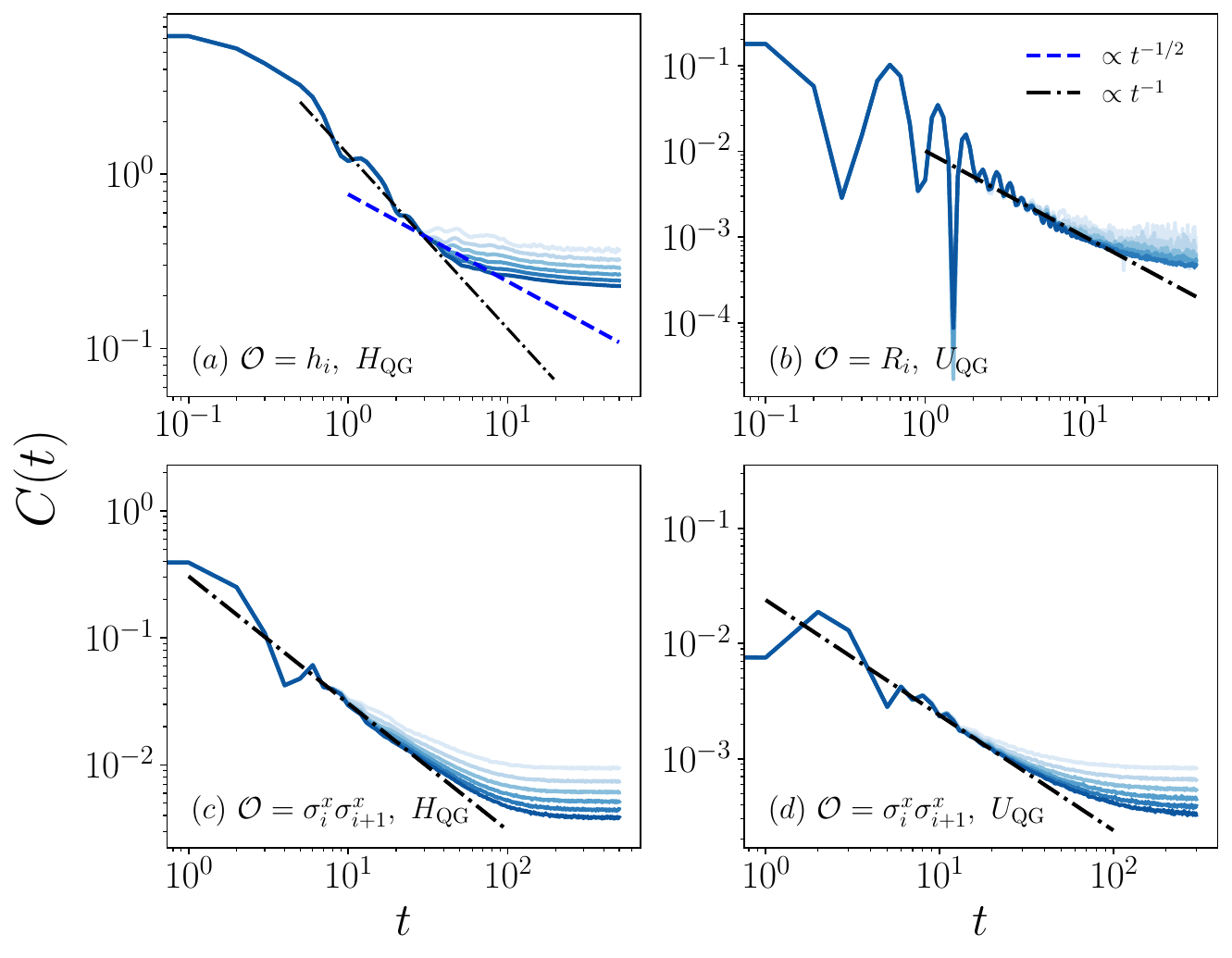}
\caption{Infinite-temperature autocorrelation functions $ C(t) = \mathrm{Tr}(\mathcal{O}(t)\mathcal{O})$ for the operators $ \mathcal{O} = h_i $ [panels (a) and (b)] and $ \mathcal{O} = \sigma_i^x \sigma_{i+1}^x $ [panels (c) and (d)] in the Hamiltonian and Floquet models, respectively, 
$ q = 2.6 $, $ \lambda = 1.0 $
and system sizes $L = 20,22,24,26,28,30$ (color from light to dark). 
Dotted ($1/t^{1/2}$) and dashed ($1/t$) lines are guides to the eye.}
\label{fig_C_RRt}
\end{figure}

Next, we consider correlation functions of other operators, including operators that have $U(1)$ charge two and zero (such as the Hamiltonian density). These should overlap with composite objects made out of hydrodynamic fields, see Refs.~\cite{Glorioso:2020loc,Matthies:2024lqx} for examples. Fig.~\ref{fig_C_RRt} shows results consistent with this expectation: these operators decay faster than longitudinal and transverse spin, but still in a power-law fashion. One of these is somewhat surprising, the decay of energy density in the Hamiltonian model shown in Fig.~\ref{fig_C_RRt}(a), which appears to be faster than diffusive. We have found that energy density correlators exhibit conventional diffusion in the classical model with QG symmetry \eqref{eq_classical_QG_model}, suggesting that the apparent superdiffusion in Fig.~\ref{fig_C_RRt}(a) may be a finite time/size effect.

%######################################################################%
%======================================================================%
%======================================================================%
%======================================================================%
%######################################################################%
\section{Transverse spin decay without QG symmetry}\label{app_stretchedexp}

\begin{figure}[t]
\includegraphics[width=1.0\columnwidth]{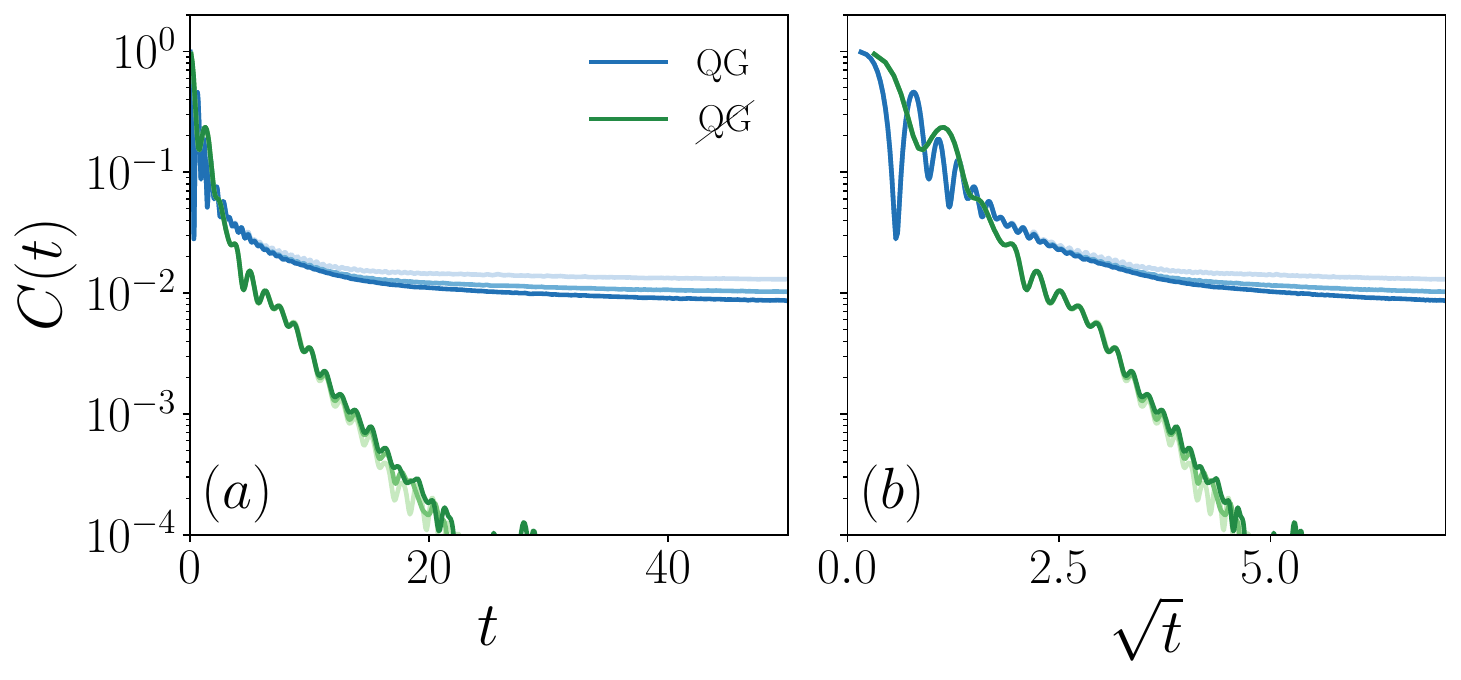}
\caption{Infinite-temperature autocorrelation functions $ C(t) = \mathrm{Tr}(\mathcal{O}(t)\mathcal{O})$ for the transverse spin operators $ \mathcal{O} = \sigma^x$ in the
XXZ model with and without QG symmetry as a function of (a) $t$ and (b) $\sqrt{t}$, 
Hamiltonian and Floquet models, respectively, for system size $L = 22,26,30$ (color from light to dark). The parameters are $ q = 2.6 $ and $ \lambda = 1.0 $.}
\label{fig_C_stretchedexp}
\end{figure}

In generic many-body systems with a $U(1)$ symmetry, operators charged under the symmetry are expected to decay very quickly because their charge forbids any overlap with composite hydrodynamic fields. One natural expectation is for this decay to be exponential, $\langle \sigma_+(t)\sigma_-\rangle \sim e^{-\Gamma t}$, as arises in holographic models and lattice models in the limit of large local Hilbert spaces. This expectation or at least its applicability to generic models was recently challenged in Refs.~\cite{McCulloch:2025fzk,McCulloch:2026pra}, which proposed that rare region effects lead to slower stretched exponentials $e^{-t^{\alpha}}$, $0<\alpha< 1$, at least in $d=1$ spatial dimension. Our results on the XXZ spin chain with next-nearest neighbor interactions and without QG symmetry, shown in Fig.~\ref{fig_C_stretchedexp}, seem consistent with this proposal. While transverse spin clearly decays much faster than in any of the models with QG symmetry, it appears to decay a little more slowly than exponential; this last point however would require further numerics to confirm. 

\begin{figure}[t]
\includegraphics[width=1\columnwidth]{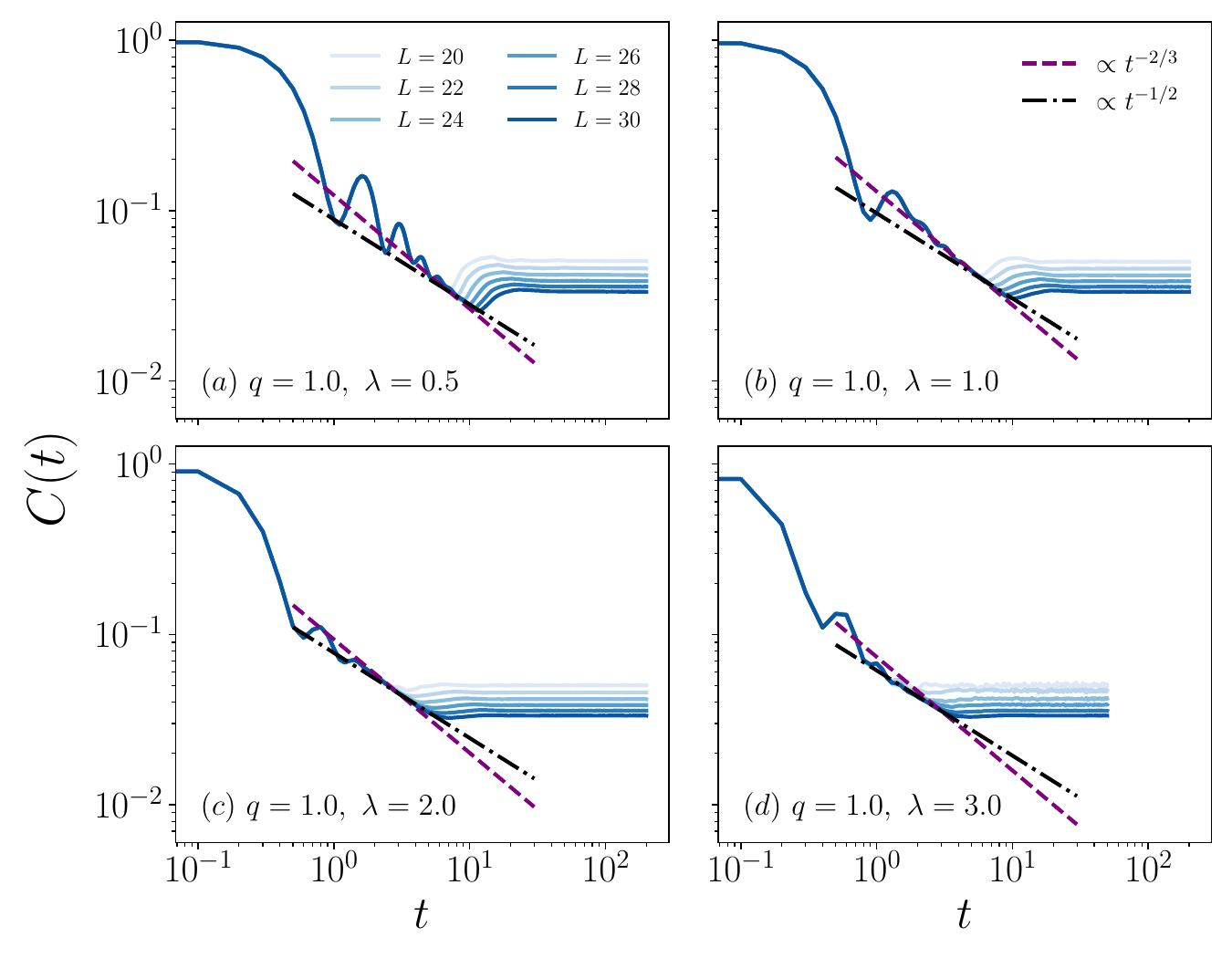}
\caption{Infinite temperature autocorrelation function $C(t)= \Tr(\mathcal O(t)\mathcal O)$ for transverse spin $\mathcal O = \sigma^x$ in the XXZ model, with $q=1$, and $\lambda =0.5,1.0,2.0,3.0$ [(a)-(d)],
for system size $L = 20,22,24,26,28,30$ (color from light to dark). Dotted ($1/t^{1/2}$) and dashed ($1/t^{2/3}$) lines are guides to the eye.}
\label{fig_c2}
\end{figure}

We also confirm conventional diffusive behavior in our chaotic model \eqref{eq_HQG} when $q=1$, in which case the QG reduces to conventional $SU(2)$. See Fig.~\ref{fig_c2}.

%######################################################################%
%======================================================================%
%======================================================================%
%======================================================================%
%######################################################################%
\section{Details about the classical model with quantum group symmetry} \label{app:classical}
We start from classical spins variables living on a sphere, $|\vec{S}_i|=1$.
The spins obey the Poisson brackets
\begin{equation}
    \{S^a_i,S^b_j\}=\epsilon^{abc} \delta_{ij}S^c_i\,,
\end{equation}
or, defining $S^\pm = S^x \pm i S^y$,
\begin{equation}
    \{S^\pm_i,S^z_j\}= \pm i \delta_{ij}S^\pm_i\,, \{S^+_i,S^-_j\}= -2i \delta_{ij}S^z_i\,.
\end{equation}
Now we define the generators
\begin{equation}
    K_j=e^{\eta S^z_j},
    \qquad
    E_j= f(S^z_j) S^+_j,
    \qquad
    F_j=-\frac{\eta}{\sinh\eta}\, f(S^z_j) S^-_j 
\end{equation}
where the function $f$ is chosen as
\begin{equation}
 f(z)
    =
    {1\over \eta}
    \left(
        {\cosh\eta -\cosh \eta z\over 1-z^2}
    \right)^{1/2}\,.
\end{equation}
This form of $f$ is chosen such that $E_j$, $F_j$ and $K_j$ satisfy the quantum group algebra with the parameter $q=e^\eta$,
\begin{equation}
\begin{gathered}
		\{K_i,E_j\}=-i\eta\, \delta_{ij} K_j E_j\, ,\qquad \{K_i,F_j\}=i\eta\, \delta_{ij} K_j F_j,\\
		\{E_i,F_j\} =  i \delta_{ij} \,{K_j-K_j^{-1}\over e^\eta-e^{-\eta}}.
\end{gathered}
\end{equation}
% \BZ{Mention that Casimir is satisfied because spins are living on a sphere}.
The Casimir of the quantum group $C_i=E_i F_i-\frac{1}{2\eta \sinh \eta}(K_i+K_i^{-1})$ is a constant because of the constraint $|\vec{S}_i|=1$, and indeed commutes with all the generators of the quantum group. We can define the generalization of these generators as acting on $N$ variables using the symmetric coproduct
\begin{equation}
\begin{gathered}
E = \sum_i \prod_{j<i} K_{j}^{-1/2} E_i \prod_{j>i}K_j^{1/2}\ \\
F = \sum_i \prod_{j<i} K_{j}^{-1/2} F_i \prod_{j>i}K_j^{1/2}\ \\
K = \prod_i K_i\,,
\end{gathered}
\end{equation}
which satisfy the same quantum group commutation relations.
Then we can construct the Hamiltonian
\begin{equation}
\begin{split}
H = - \sum_{i=1}^{L-1}\bigg[
&K_i^{-1/2}
\left(
E_{i+1}F_i
+
E_iF_{i+1}
\right)
K_{i+1}^{1/2}
\\
&-
\frac{1}{\eta \sinh \eta} K_i^{-1}K_{i+1}
+ \frac{1}{\eta \tanh \eta} (K_i^{-1}+K_{i+1})
%\left(K_i^{-1}K_{i+1} -\cosh\eta\, K_{i+1} -\cosh\eta\, K_i^{-1}\right)
\bigg]\,.
\end{split}
\end{equation}
This is quantum group symmetric, i.e. $\{E,H\}=\{F,H\}=\{K,H\}=0$.

To time evolve the system and obtain correlation functions such as $\langle S^x_a(t) S^x_a(0) \rangle$, we numerically integrate the equations 
\begin{equation}
    \frac{d}{dt}S^a_i = \{ S^a_i,H \}\,.
\end{equation}
In practice, we parametrize the spins by the two polar angles and evolve them using a fourth-order Runge-Kutta algorithm. We start from a random configuration of spins and compute correlation functions by averaging over many different samples and different spins in a region around the center of the chain.

%######################################################################%
%======================================================================%
%======================================================================%
%======================================================================%
%######################################################################%
\section{Finite size effects in hydrodynamics}\label{app_finitesize}

Finite size scaling collapse of autocorrelation functions is a computationally cheap approach to determining the dynamic critical exponent $z$. This approach does {\em not} apply to operators protected by QG symmetry, due to subtle finite size effects discussed in Sec.~\ref{sec_finite_size}. Nevertheless, this technique may be useful in studying anomalous hydrodynamics more generally. 

\begin{figure}[t]
\includegraphics[width=1\columnwidth]{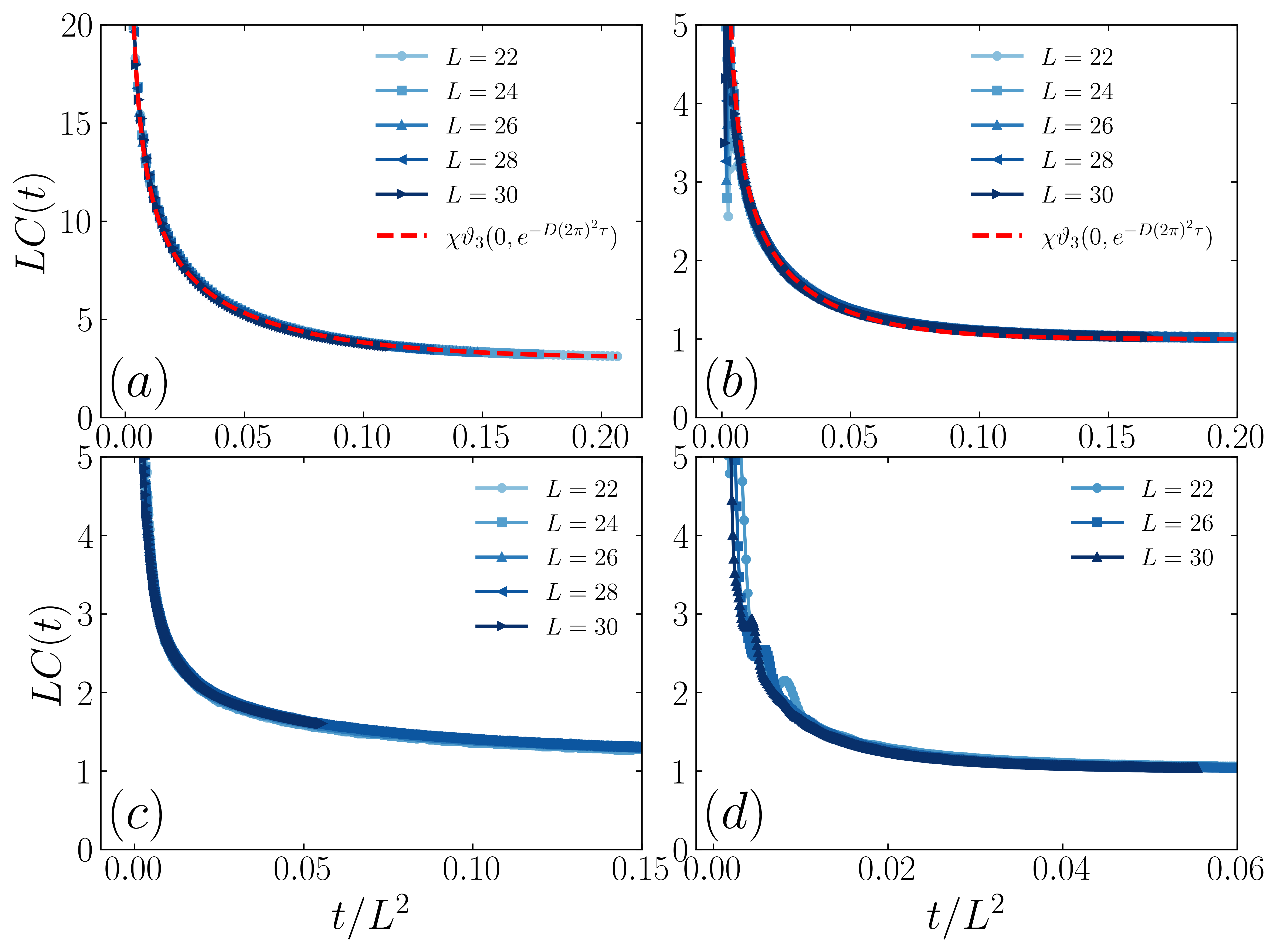}
\caption{Scaling collapse of $L C(t)$ versus $t/L^2$ for (a) energy-density correlations, ${\cal O}=h$, in the mixed-field Ising model and (b) longitudinal-spin correlations, ${\cal O}=\sigma^z$, in the chaotic XXZ model, compared with the theoretical prediction in Eq.~\eqref{eq_Fscaling_finitesize} (dashed lines). Corresponding collapses in two QG-symmetric models are shown in (c,d).}
\label{fig_e1}
\end{figure}

As described in Sec.~\ref{sec_finite_size}, correlation functions of conserved densities in a finite volume are universal properties of the dissipative universality class and the boundary conditions. For example, Eq.~\eqref{eq_Fscaling_finitesize} shows the prediction for diffusion with periodic boundary conditions. This scaling function is compared to the autocorrelation function of diffusing lattice models in Fig.~\ref{fig_e1}(a) and (b). The example of energy diffusion in the mixed field Ising model, \ref{fig_e1}(a), is particularly striking: the larger system size data is almost entirely redundant, as it essentially exactly collapses to the smaller system size data. In the chaotic XXZ model, \ref{fig_e1}(b), small residual corrections to scaling are still visible; these should be due to intermediate time power-law corrections to hydrodynamics.

The boundary conditions are more subtle in the QG models we study. It is unclear what boundary condition on the hydrodynamic EFT descends from the specific QG-preserving boundary terms of the open spin chain in \eqref{eq_XXZ} and \eqref{eq_HQG}. However, even when the precise boundary conditions are not known, finite size scaling collapse can be used to find the dynamic critical exponent: one searches for collapse of the data $L\langle n(t)n\rangle$ as a function of $t/L^z$. Figs.~\ref{fig_e1}(c) and (d) show that longitudinal spin correlators collapse near the diffusive value $z=2$.

Compared to the conventional approach of extracting $z$ from the autocorrelation function at intermediate times before finite size effects arise, the approach described here uses all the late time data including those with large finite size effects. It is thus less affected by the earlier time data which suffers from power-law corrections to diffusion \eqref{eq_nn}.

%######################################################################%
%======================================================================%
%======================================================================%
%======================================================================%
%######################################################################%
\section{Mazur bound for QG charge}\label{app_Mazur}

The Mazur bound \eqref{eq:Mazur1} can be generalized to the presence of several charges as
\begin{equation} \label{eq:Mazur3}
C(\infty) = \lim_{T\to \infty}\frac1{T}\int_0^T dt \, \langle \mathcal{O}^\dagger(t) \mathcal{O}\rangle 
	\geq \langle \mathcal{O}^\dagger Q_a\rangle(G^{-1})_{ab}\langle Q_b^\dagger\mathcal{O} \rangle
\end{equation}
where $G$ is the Gram matrix
\begin{equation}
    G_{ab}  = \langle Q_a^\dagger Q_b \rangle\,.
\end{equation}

While in \eqref{eq:Mazur1} we only included $E$, we will now see that including other conserved charges (such as $E^2 F$ or $q^H E$) will improve the bound.

\begin{figure}[t]
    \includegraphics[width=0.9\linewidth]{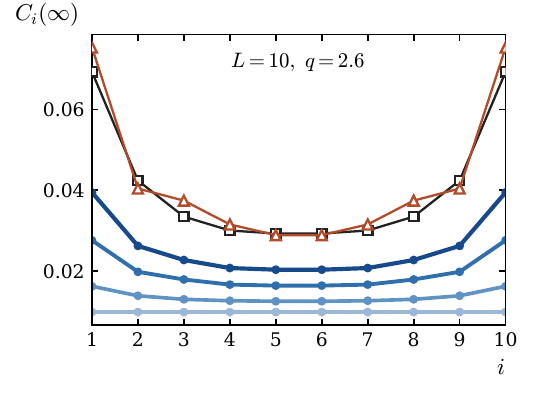}
    \caption{Improved Mazur bound and exact results for $C_i(\infty) = \lim_{T\to \infty}\frac1{T}\int_0^T dt\langle \sigma^+_{i}(t)\sigma^-_{i} \rangle $ for a spin chain with $L=10$, $q=2.6$. In blue we show lower bounds obtained with an increasing number of operators considered, from 1 (light blue) to 13 (dark blue); we also show the exact $C_i({\infty})$ values for $\lambda=0$ (black) and $\lambda=1$ (red).}
    \label{fig:Mazur}
\end{figure}

We also mention that the choice of coproduct plays a role in equation \eqref{eq:Mazur3}. This follows from the fact different coproducts imply different conjugation rules.
The symmetric coproduct %\BZ{Check i'm using the right one}\AZ{Change $q \to q^{-1}$?}
\begin{equation}
\begin{aligned}
    \Delta(E) &= E \otimes q^{H/2} + q^{-H/2}\otimes E \\
    \Delta(F) &= F \otimes q^{H/2} + q^{-H/2}\otimes F \\
\end{aligned}
\end{equation}
is consistent with the conjugation relation
\begin{equation}
    E^\dagger = F\,.
\end{equation}
Instead, consistency of a different coproduct such as
\begin{equation}
\begin{aligned}
    \Delta(E) &= E \otimes \mathds{1}+ q^{-H}\otimes E \\
    \Delta(F) &= F \otimes q^{H}+ \mathds{1}\otimes F \\
\end{aligned}
\end{equation}
requires the conjugation relation
\begin{equation}
    E^\dagger = q^{-H}F\,.
\end{equation}
We choose the symmetric coproduct for which the conjugation relations are simpler, $E^\dag=F$. Considering \eqref{eq:Mazur3} for the same set of charges but with different choices of coproduct will give different Mazur bounds.

Considering only the operator $E$, we find the bound \eqref{eq:Mazur2}. 
We now include more and more conserved charges, such as $q^H E$, $E^2F$, $q^H E^2F$ and so on. In the $SU(2)$ case, adding these operators would not change the bound; however, for $q\neq 1$, it improves as we add more charges. Now the bound depends on the position of the spin we are considering, and we show some examples in Figure \ref{fig:Mazur} for a spin chain of length $L=10$. We have checked explicitly for small values of $L$ that when including enough charges, the bound becomes insensitive to the choice of coproduct.

\section{Detailed gate ordering of the Floquet model}
\label{app:floquet_map}

In this section, we specify the precise Floquet operator used in the numerical simulations. 
We label sites by \(0,1,\ldots,L-1\). A two-site gate
\(R^{\mathrm{QG}}_j\) acts on sites \((j,j+1)\), while a three-site gate
\(V^{\mathrm{QG}}_j\) acts on sites \((j,j+1,j+2)\). The Floquet operator for one period reads
\begin{equation}
U=U_3U_2 .
\end{equation}

The two-site layer is
\begin{equation}
U_2 =
\prod_{j}
e^{-i\tau R^{\mathrm{QG}}_{i^{(2)}_1+2\ell}}
\prod_{j}
e^{-i\tau R^{\mathrm{QG}}_{i^{(2)}_0+2\ell}},
\end{equation}
where
\begin{equation}
i^{(2)}_0=\frac{L}{2}\bmod 2,
\qquad
i^{(2)}_1=(i^{(2)}_0+1)\bmod 2 .
\end{equation}

The three-site layer is
\begin{equation}
U_3 =
\prod_{\ell}
e^{-i\tau\lambda V^{\mathrm{QG}}_{i^{(3)}_2+3\ell}}
\prod_{j}
e^{-i\tau\lambda V^{\mathrm{QG}}_{i^{(3)}_1+3\ell}}
\prod_{\ell}
e^{-i\tau\lambda V^{\mathrm{QG}}_{i^{(3)}_0+3\ell}},
\end{equation}
with
\begin{equation}
i^{(3)}_0=\frac{L}{2}\bmod 3,
\qquad
i^{(3)}_1=(i^{(3)}_0+1)\bmod 3,
\end{equation}
and
\begin{equation}
i^{(3)}_2=(i^{(3)}_0+2)\bmod 3 .
\end{equation}
In the products above, \(j\) runs over all non-negative integers such that the
corresponding gate is fully contained in the chain.

The offsets are chosen to pin the brick-wall pattern to the center of the
chain. With this convention, the local gate structure around the center is
kept fixed as \(L\) is varied, which minimizes the system-size dependence of
the short-time dynamics of autocorrelation functions of local operators. The precise staggered Floquet operator is also defined similarly.

\vfill

\bibliography{qg_hydro}

\end{document}